\PassOptionsToPackage{pdfpagelabels=false}{hyperref}
\PassOptionsToPackage{nonamebreak}{natbib}   
\documentclass[fleqn,usenatbib]{mnras}
\usepackage{newtxtext,newtxmath}

\usepackage[T1]{fontenc}
\usepackage{ae,aecompl}
\pdfoutput=1

\usepackage{graphicx}	
\usepackage{amsmath}	
\usepackage{amssymb}	
\usepackage{subfigure}

\newcommand{\FIREurl}{\href{http://fire.northwestern.edu}
{\url{http://fire.northwestern.edu}}}
\newcommand{\gizmourl}{\href{http://www.tapir.caltech.edu/~phopkins/Site/GIZMO.html}{\url{http://www.tapir.caltech.edu/~phopkins/Site/GIZMO.html}}}
\newcommand\altaffilmark[1]{$^{#1}$}
\newcommand\altaffiltext[1]{$^{#1}$}

\title[In search of the most ancient stars]{Where are the most ancient stars in the Milky Way?}

\author[El-Badry et al.]{
\parbox[t]{\textwidth}{ 
Kareem El-Badry\thanks{E-mail: kelbadry@berkeley.edu}\altaffilmark{1},
Joss Bland-Hawthorn\altaffilmark{2, 3, 4},
Andrew Wetzel\altaffilmark{5}, 
Eliot Quataert\altaffilmark{1},
Daniel R. Weisz\altaffilmark{1},
Michael Boylan-Kolchin\altaffilmark{6},
Philip F.~Hopkins\altaffilmark{7},
Claude-Andr{\'e} Faucher-Gigu{\`e}re\altaffilmark{8},
Du\v{s}an Kere\v{s}\altaffilmark{9}, 
Shea Garrison-Kimmel\altaffilmark{7}
} 
\vspace*{6pt} \\
\altaffiltext{1}{Department of Astronomy and Theoretical Astrophysics Center, University of California Berkeley, Berkeley, CA 94720} \\
\altaffiltext{2}{Miller Professor, Miller Institute, University of California Berkeley, Berkeley, CA 94720, USA} \\
\altaffiltext{3}{Centre of Excellence for All Sky Astrophysics in 3 Dimensions (ASTRO-3D), Australia} \\
\altaffiltext{4}{Sydney Institute for Astronomy, School of Physics, University of Sydney, NSW 2006, Australia} \\
\altaffiltext{5}{Department of Physics, University of California, Davis, CA 95616} \\
\altaffiltext{6}{Department of Astronomy, The University of Texas at Austin, Austin, TX 78712} \\
\altaffiltext{7}{TAPIR, Mailcode 350-17, California Institute of Technology, Pasadena, CA 91125} \\
\altaffiltext{8}{Department of Physics and Astronomy and CIERA, Northwestern University, Evanston, IL 60208} \\ 
\altaffiltext{9}{Department of Physics, Center for Astrophysics and Space Sciences, University of California at San Diego, La Jolla, CA 92093} \\ 
}

\date{Accepted to MNRAS}

\pubyear{2018}

\begin{document}
\label{firstpage}
\pagerange{\pageref{firstpage}--\pageref{lastpage}}
\maketitle

\begin{abstract}
The oldest stars in the Milky Way (MW) bear imprints of the Galaxy's early assembly history. We use FIRE cosmological zoom-in simulations of three MW-mass disk galaxies to study the spatial distribution, chemistry, and kinematics  of the oldest surviving stars ($z_{\rm form} \gtrsim 5$) in MW-like galaxies. 
We predict the oldest stars to be less centrally concentrated at $z=0$ than stars formed at later times as a result of two processes. 
First, the majority of the oldest stars are not formed {\it in situ} but are accreted during hierarchical assembly. These {\it ex situ} stars are deposited on dispersion-supported, halo-like orbits but dominate over old stars formed {\it in situ} in the solar neighborhood, and in some simulations, even in the galactic center. 
Secondly, old stars formed {\it in situ} are driven outwards by bursty star formation and energetic feedback processes that create a time-varying gravitational potential at $z\gtrsim 2$, similar to the process that creates dark matter cores and expands stellar orbits in bursty dwarf galaxies. 
The total fraction of stars that are ancient is more than an order of magnitude higher for sight lines {\it away} from the bulge and inner halo than for inward-looking sight lines.
Although the task of identifying specific stars as ancient remains challenging, we anticipate that million-star spectral surveys and photometric surveys targeting metal-poor stars already include hundreds of stars formed before $z=5$. We predict most of these targets to have higher metallicity ($-3 < \rm [Fe/H] < -2$) than the most extreme metal-poor stars.
\end{abstract}



\section{Introduction} 
\label{sec:intro}

The oldest stars in the Milky Way are relics of star formation in the early Universe, providing a probe of physical processes that can otherwise be studied only at high redshift \citep[e.g.][]{Freeman_2002, Beers_2005, Frebel_2015}. Determining the properties, phase-space distribution, and stellar yields of first- and second-generation stars is therefore a primary goal of near-field cosmology \citep[e.g.][]{Abel_2000, Bromm_2004, Karlsson_2013, Frebel_2015}. Ongoing  spectroscopic and photometric surveys of the Galaxy have already begun to identify large numbers of metal-poor stars suspected to be ancient \citep[e.g.][]{Beers_1985, Christlieb_2003, Helmi_2003, Yong_2013, GarciaPerez_2013, Schlaufman_2014, Casey_2015, Howes_2015, Howes_2016, Li_2015, Minniti_2016, Starkenburg_2017b, Cescutti_2017}, and parallel efforts to measure precise atmospheric parameters, masses, ages, and detailed abundance patterns for large samples of these stars are underway \citep[e.g.][]{Rauer_2014, Ricker_2015, Li_2015, Feltzing_2017, FernandezAlvar_2017, Sharma_2018}. 

Given the archaeological interest in identifying very old stars, there arise the questions of (a) where, in terms of both physical location and metallicity, ancient stars can most efficiently be found, and (b) how metallicity and age are correlated for old stars. Numerous studies have shown that the MW's stellar halo and satellite galaxies retain a wealth of information about the Galaxy's assembly history and earliest stellar populations \citep[e.g.][]{Bullock_2001, Bullock_2005, Salvadori_2007, Helmi_2008, Kirby_2008, Bovill_2011, Brown_2014, Magg_2018, Beniamini_2018}. The observational fact that the stellar halo consists primarily of metal-poor stars broadly supports this notion. On the other hand, cold dark matter (CDM) simulations predict that the number density of old stars should be highest near the Galactic center, in the bulge and innermost stellar halo ($r\lesssim  3$\,kpc; e.g. \citealt{White_2000, Brook_2007, Tumlinson_2010, Gao_2010, Salvadori_2010, Starkenburg_2017, Griffen_2018}. Due to efficient enrichment in regions of high SFR density at early times, ancient stars found in the inner Galaxy are predicted to be more metal-rich than those in the outer halo ($r\gtrsim 20$\,kpc). 

The search for ancient and metal-poor stars has a long observational history.\footnote{For a full history, see \citet{Sandage_1986}, and references therein.} Metal-poor RR Lyrae stars were identified in the bulge and inner stellar halo nearly a century ago \citep{Baade_1946, Baade_1951} and were subsequently associated with first- and/or second-generation stars \citep{Walker_1991, Soszynski_2011, Minniti_2016}. The kinematics of metal-poor halo stars were used by \citet{Eggen_1962} in an early attempt to constrain the Milky Way's formation history. \citet{Wallerstein_1963} proposed to use the abundance patterns of metal-poor halo giants to constraint Galactic chemical enrichment models.

More recently, the search for ancient stars has targeted ultra metal-poor (UMP) stars ($[\rm Fe/H] \lesssim -4$; e.g., \citealt{Keller_2014, Aguado_2018}) and somewhat higher-metallicity stars ($\rm [Fe/H] < -2$) with $r$-process enhancement. UMP stars are of particular interest because they are diagnostic of the yields of first-generation stars \citep{Beers_2005, Karlsson_2013, Norris_2013, Bessell_2015}; in some cases, their abundance patterns appear consistent with enrichment by a single supernova. However, they are quite rare: only a few dozen stars have been identified with $[\rm Fe/H] < -4$ \citep{Frebel_2015, Starkenburg_2017}, making a statistical study of the population challenging. It also remains unclear whether UMP stars are unambiguously ancient, as many theoretical models predict them to continue forming in low-density environments until relatively late times (until $z= 2-3$; \citealt{White_2000, Brook_2007, Tornatore_2007, Trenti_2009}). 

Metal-poor stars with enhanced $r$-process elements and no detectable or weak $s$-process enrichment are thought to have formed at an early time before the onset of the AGB phase (likely $\lesssim 300$ Myr after the formation of the first stars; \citealt{Hill_2002, Frebel_2015, Ji_2016, Hansen_2017, Ji_2018}). A few such stars have measured radioactive lifetimes that indicate that they are very old \citep{Sneden_1996, Cayrel_2001, Frebel_2007,Hill_2017}, albeit with systematic uncertainties. $r$-process enhanced metal-poor stars have been identified in the bulge, in the stellar halo, and in Local Group dwarf galaxies. 

\begin{table*}
\centering
\caption{Summary of simulations}
\label{tab:properties}
\begin{tabular}{p{0.8cm} | p{1.5cm}| p{1.5cm} |  p{1.5cm} | p{1.5 cm} | p{1.8 cm} | p{0.8cm} | p{0.8cm} } 

Name &  $\log (M_{\rm star})$  $(\rm M_{\odot})$ & $\log (M_{200 \rm m})$ $(\rm M_{\odot})$ & $f_{{z_{\rm form}}>5}$ & $f_{{\rm \left[Fe/H\right]}<-2}$ & $f_{{z_{\rm form}}>5},\,|$ $\rm \left[Fe/H\right]<-2$ & $m_{\rm b}$ $(\rm M_{\odot})$ & $m_{\rm DM}$ $(\rm M_{\odot})$ \\
\hline
\texttt{m12i}   &  10.8 &  12.1   &   0.0009  & 0.0063  & 0.12  & 7070  & 35200 \\
\texttt{m12f}   &  10.9 &  12.2   &   0.0029  & 0.0031  & 0.49  & 7070  & 35200 \\
\texttt{m12m}   &  11.1 &  12.2   &   0.0011  & 0.0031  & 0.26  & 7070  & 35200 \\

\hline
\end{tabular}
\begin{flushleft}
$M_{\rm star}$ is the stellar mass within $3\times R_{1/2}$, where $R_{1/2}$ is the 3D stellar half-mass radius. $M_{200 \rm m}$ is the total mass within $R_{\rm 200m}$, where $R_{200 \rm m}$ is the radius within which the matter density is $200 \times$ the mean matter density. $M_{\rm star}$ and $M_{\rm 200m}$ ar reported at $z=0$. $f_{z_{{\rm form}}>5}$ and $f_{{\rm \left[Fe/H\right]}<-2}$ are the fraction of stars within 10 kpc at $z=0$ that formed before $z=5$ and have $[\rm Fe/H] < -2$, respectively. $f_{z_{{\rm form}}>5|{\rm \left[Fe/H\right]}<-2}$ is the fraction of stars with $[\rm Fe/H] < -2$ that formed before $z=5$. 
The last two columns show $m_{\rm b}$ and $m_{\rm DM}$, the average baryon and dark matter particle masses. 
\end{flushleft}
\end{table*}

In this work, we study the oldest stars in {\it simulated} galaxies with the goals of making predictions for MW  surveys and better understanding the origin of the oldest stars already being observed. We focus on three $L_\star$ disk galaxies from the FIRE project\footnote{See the FIRE project website at \FIREurl} with $z=0$ observable properties that are broadly consistent with the MW. We first present the properties of the oldest stars at $z=0$ and then trace the stars back to their formation sites, quantifying the effects of accretion stars formed {\it ex situ} and outward migration of stars formed {\it in situ} on the distribution of all old stars at $z=0$. 

Many previous theoretical works \citep[e.g.][]{Scannapieco_2006, Salvadori_2007, Salvadori_2010, Tumlinson_2010, Gao_2010, Komiya_2010, Ishiyama_2016, Griffen_2018} have predicted the $z=0$ spatial distribution of old stars in MW-like galaxies using simple analytic prescriptions for the formation sites of ancient stars combined with dark-matter-only simulations or Monte-Carlo merger trees  based on Press-Schechter theory. These works have all predicted old stars to be centrally concentrated at $z=0$ because the earliest stellar generations are predicted to form in the highest density peaks, and these preferentially end up near the center of the primary halo by $z=0$; i.e, in the bulge. Some studies \citep{Brook_2007, Starkenburg_2017, Sharma_2017} have also arrived at similar conclusions using cosmological hydrodynamics simulations including star formation and cooling. 

In this work, we study the formation and subsequent evolution of ancient stars in cosmological zoom-in simulations that (a) self-consistently include the effects of baryonic feedback, (b) explicitly model a multiphase ISM, leading to bursty star formation at high redshift and in low-mass halos, and (c) produce realistic MW-like galaxies at $z=0$. 
As we will show, baryonic feedback processes arising from bursty star formation have non-negligible effects on the late-time distribution of old stars. In particular, we find that feedback-driven fluctuations in the gravitational potential at high redshift drive old stars that are formed near the galactic center outward, into the outer bulge and inner stellar halo. This effect has not been captured by previous studies of ancient stars, which have either ignored the effects of baryons altogether or have adopted a simplified model of the ISM that suppresses the burstiness of star formation (see further discussion in Section~\ref{sec:sim_compare}).

The remainder of this paper is organized as follows. 
In Section~\ref{sec:sim_description}, we introduce the FIRE simulations.
In Section~\ref{sec:aitoff}, we show the predicted spatial, chemical, and kinematic distributions of the oldest stars at $z=0$. 
In Section~\ref{sec:where_form}, we wind back the clock and identify the formation sites for the oldest stars. 
In Section~\ref{sec:sim_compare}, we compare our results to earlier work. 
We summarize our findings in Section~\ref{sec:discussion} and discuss prospects for identifying the surviving population of ancient stars in ongoing MW surveys. In Appendix~\ref{sec:diff_coeff}, we examine how sensitive our results are to changes in mass resolution and in the turbulent diffusion coefficient.

\section{FIRE simulations}
\label{sec:sim_description}

We study cosmological zoom-in simulations of three MW-mass galaxies from the FIRE project \citep{Hopkins_2014}. The simulations were run with the \texttt{GIZMO}\footnote{A public version of the \texttt{GIZMO} code is available at \gizmourl.} hydrodynamics code \citep{Hopkins_2015} in the Lagrangian ``meshless finite mass'' (MFM) mode, using the FIRE-2 model for galaxy formation and feedback \citep{Hopkins_2017}. For details regarding the physical processes modeled in these simulations and their numerical implementation, we refer to \citet{Hopkins_2017} and \citet{Hopkins_2017b}. 

The three halos we study were first  presented by \citet{Wetzel_2016} and \citet{Hopkins_2017}. Their properties are summarized in Table~\ref{tab:properties}. At $z=0$, they host galaxies with structural parameters broadly similar to the MW. They have realistic stellar disks and bulge-to-disk ratios \citep{ElBadry_2018, GarrisonKimmel_2017}, HI rotation curves and velocity dispersions \citep{ElBadry_2018b}, satellite populations (\citealt{Wetzel_2016}; Garrison-Kimmel et al., in prep.),  and stellar halos \citep{Bonaca_2017, Sanderson_2017}. 

Although the three simulated galaxies have similar gross structural properties at $z=0$, we will show in Section~\ref{sec:where_form} that their early assembly histories differ from one another substantially, as is common in CDM \citep{Cooper_2010}. Given that we do not currently identify a clear reason to prefer one simulated MW-analog over another we view the three simulations as realizations of plausible assembly histories that could produce a MW-like galaxy at $z=0$. Without firmer priors on the formation history of the MW, the scatter between the simulations sets a lower limit on the uncertainty of our predictions as applied to the real MW. 

Due to significant uncertainties in the properties and formation process of zero-metallicity stars \citep[see e.g.][]{Bromm_2004, McKee_2008, Wise_2012,  Bromm_2013RPPh}, the FIRE model does not attempt to explicitly model Pop III stars. Instead, the abundances of all baryon particles in the simulation are set to a metallicity floor of $[{\rm M}_{i}/{\rm H}] = -4$ in the initial conditions, where $[{\rm M}_{i}/{\rm H}]$ denotes the logarithmic abundance of individual metal species compared to their Solar value. The specific value of the metallicity floor is somewhat arbitrary; a floor is required to prevent numerical problems in cooling, and $[{\rm M}_{i}/{\rm H}] = -4$ is similar to the typical value expected after enrichment by Pop III stars \citep{Bromm_2011, BlandHawthorn_2015}. Different metallicity floors have been used elsewhere in the literature \citep[e.g.][]{Starkenburg_2017}, but systematic studies of the effects of varying the metallicity floor on the formation sites of ancient stars have yet to be carried out and represent a promising avenue for future work. The first generation of stars formed in the simulation quickly enrich the surrounding gas, so that most stars formed after $z \sim 10$ are enriched to $[{\rm M}_{i}/{\rm H}] \sim -3$, a factor of 10 enhancement over the metallicity floor. The metallicity floor is therefore not expected to significantly bias the abundances of these stars; nevertheless, we caution against over-interpreting the abundances of the most metal-poor stars in the simulation $[{\rm M}_{i}/{\rm H}] \lesssim -3.5$. 

The realizations of the simulations studied in this work, unlike the first realization presented in \citet{Wetzel_2016}, were run with the subgrid model for turbulent diffusion of metals described in \citet{Hopkins_2017}, using a diffusion coefficient $C_{0}\approx 0.003$. Although the inclusion of turbulent metal diffusion has a negligible effect on galaxies' overall structural properties, including total metallicity \citep{Su_2017, Hopkins_2017}, it has been shown to improve metal mixing in the ISM and produce more realistic metallicity distribution functions, particularly preventing the unphysical formation of very low-metallicity stars at late times (\citealt{Escala_2018}; Wetzel et al., in prep). We explore the model's sensitivity to the choice of diffusion coefficient in Appendix~\ref{sec:diff_coeff}. There we show that our results are relatively insensitive to the choice of diffusion coefficient, though the properties of the most ancient stars ($z_{\rm form} \gtrsim 7$) do vary somewhat with simulation resolution.

Each star particle in the simulation represents a simple stellar population with fixed abundances, uniform age, and a typical initial mass of 7070 M$_{\odot}$. For a metal-poor, 13 Gyr-old stellar population with a \citet{Kroupa_2001} IMF, every 1000 M$_{\odot}$ of $z=0$ stellar mass in old star particles represents roughly 15 giant and subgiant stars (each with $m \approx 0.8 M_{\odot}$), which are the primary targets of spectroscopic surveys of the MW.\footnote{Because their ages are easier to constrain, the number of RR Lyrae stars is also potentially of interest. The frequency of RR Lyrae stars and its dependence on metallicity is poorly constrained, but a rough estimate based on measurements of globular clusters and nearby dwarf galaxies \citep{Sherwood_1975, Harris_1996, Baker_2015} is 1 RR Lyrae star per $10^4 M_{\odot}$ in $z=0$ old ($\gtrsim 10$\,Gyr) stellar mass; i.e., 1 ancient RR Lyrae for every 150 ancient giants and subgiants.}

\section{Old stars at late times}
\label{sec:aitoff}

\subsection{Spatial distribution}
\begin{figure*}
\includegraphics[width=\textwidth]{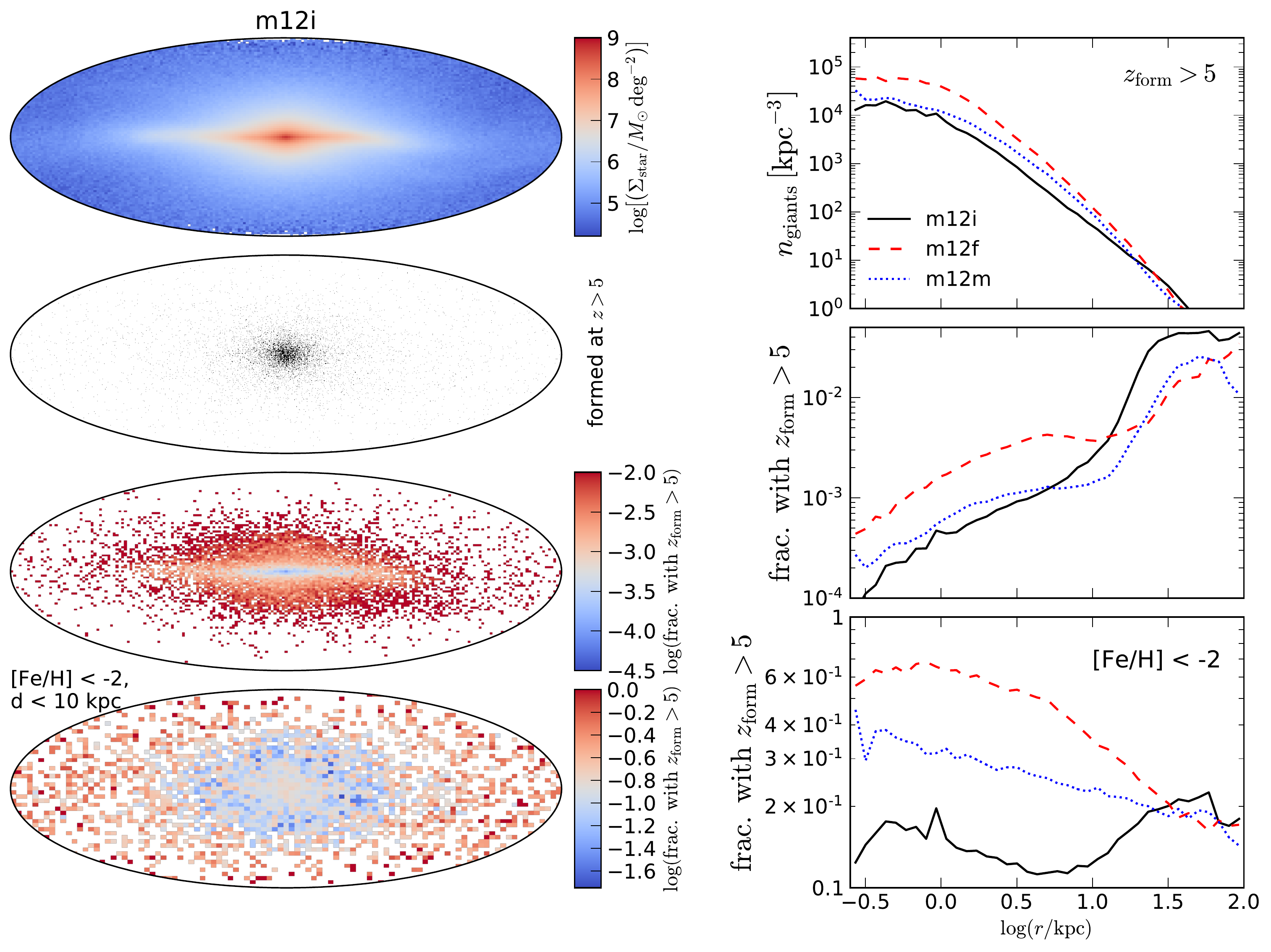}
\caption{{\bf Left}: Aitoff projections of the \texttt{m12i} simulation for an observer in the disk midplane at the solar circle ($r = 8.2$\,kpc). We show the integrated surface density of all stars today (top) and the spatial distribution of stars formed before $z=5$ (2nd panel). Because the oldest stars are distributed in a uniform spheroid extending to large radius, while later-forming stars are concentrated in the disk and central bulge, the fraction of all stars along a given line of sight that are old is \textit{lowest} for sightlines toward the galactic center (3rd panel). Bottom panel show fraction of stars with $\rm [Fe/H] <-2$ within 10\,kpc of the solar neighborbood that are ancient. {\bf Right}: Giant number density (top) and total mass fraction (middle) of old stars in three simulated MW-like galaxies. In all three galaxies, the total number density of old stars is highest near the galactic center, but old stars are less centrally-concentrated than stars formed at later times, so old stars make up a larger fraction of the total population at large radii, in the stellar halo. Bottom panel shows the fraction of metal-poor stars that are ancient; this varies only weakly with radius but is generally highest in the inner galaxy.}
\label{fig:aitoff_m12i}
\end{figure*}

Figure~\ref{fig:aitoff_m12i} shows the spatial distribution and fractional contribution to the total stellar mass of old stars in our simulations. The left panels show projections of a single galaxy, \texttt{m12i}, as viewed by an observer in the solar circle; i.e, in the disk midplane, 8.2\,kpc from the galaxy's center. In the top three panels, we include all stars within a distance $r<30$\,kpc of the galaxy's center; in the bottom panel, we consider a survey centered on the solar neighborhood and extending to a distance of 10\,kpc. 

In the second left panel of  Figure~\ref{fig:aitoff_m12i}, we show the projected sky distribution for stars that were born before $z=5$ (1.2\,Gyr after the Big Bang in our adopted cosmology), which we refer to as ``old'' or ``ancient'' stars.\footnote{We note that the cut of $z_{\rm form} > 5$ primarily yields stars younger than the oldest ``population 2.9'' stars sought by UMP star surveys, some of which may have formed as early as $z=20-30$ \citep{Frebel_2015}; however, we find no significant differences between the spatial distribution and kinematics of the first stars formed in the simulation at $z=15-20$ and those forming at $z=5$.} Here we simply plot each old star particle as a single pixel, but we remind the reader that each particle represents an entire stellar population with initial mass $\sim$7000\,M$_{\odot}$, containing of order $100$ giant and subgiant stars. Consistent with expectations from previous work \citep[e.g.][]{White_2000, Brook_2007, BlandHawthorn_2006}, this panel shows that the absolute number of old stars found in a particular line-of-sight is largest toward the dense inner regions of the galaxy. However, the total surface density of all stars -- most of which formed at later times -- is also highest toward the inner galaxy. The third panel on the left side of Figure~\ref{fig:aitoff_m12i} shows that as a result, the fraction of ancient stars increases substantially as a function of radius, particularly for sight lines out of the galactic plane. 

In the bottom-left panel of Figure~\ref{fig:aitoff_m12i}, we predict the returns of a bright magnitude-limited survey targeting metal-poor stars. We consider only star particles within 10\,kpc of the solar neighborhood, which we place at an arbitrary azimuth on a circle of radius $r=8.2$\,kpc in the disk midplane. 10 kpc corresponds roughly to the maximum distance at which an old, metal-poor giant can be detected by current MW spectroscopic surveys (e.g., GALAH or APOGEE, with magnitude limits of V$\approx$14; \citealt{DeSilva_2015, Majewski_2017}).\footnote{Our result are not sensitive to the distance limit of 10\,kpc; increasing it slightly lowers the fraction of metal-poor stars that are ancient, as the outer halo contains more metal-poor stars formed at later times.} We then show the fraction of all stars with $\rm [Fe/H] < -2$ along a given sight line that are old ($z_{\rm form} > 5$). Considering only metal-poor\footnote{Unless otherwise stated, we use ``metallicity'' to refer specifically to [Fe/H]. For the Solar abundances, we adopt $12+\log\left(n_{{\rm Fe}}/n_{{\rm H}}\right)=7.50$ and (later in the text) $12+\log\left(n_{{\rm Mg}}/n_{{\rm H}}\right) = 7.60$ from \citet{Asplund_2009}, where $n_{\rm X}$ denote number densities.}  stars removes most of the substructure due to the bulge and disk seen in the upper panels, as most bulge and disk stars in the simulated galaxies formed later ($z\gtrsim 1-2$) and have higher metallicities \citep{Ma_2017}. The bottom left panel shows that in \texttt{m12i}, the total fraction of stars that are ancient is lower toward the galactic center even when metal-rich stars are discarded. However, the opposite is true in \texttt{m12f} and \texttt{m12m} (bottom right panel).

In the right panels of Figure~\ref{fig:aitoff_m12i}, we show the 3D absolute number density and fractional mass contribution of ancient stars for all three simulated galaxies. In the top panel, we approximate the number of red giant stars contributed by every old star particle; here we assume a \citet{Kroupa_2001} IMF and use the \texttt{MIST} isochrones \citep{Choi_2016} for old, metal poor stars that were used by \citet{ElBadry_2017b}. In the middle and bottom panels, we plot the fraction of all stellar mass (middle) and of star particles with $\rm [Fe/H] < -2$ (bottom) that are ancient. Two qualitative trends in the distribution of old stars are similar for the three simulations: their absolute number density is highest near the galactic center, while the fraction of all stars that are old is highest in the outer halo. In detail, there are nontrivial differences between the simulations. The total number of old stars is higher by a factor of $\sim$5 in \texttt{m12f} than in \texttt{m12i}; despite this, the fraction of stars in the outer halo that are old is higher in \texttt{m12i}. When only iron-poor star particles are considered, the fraction of stars that are ancient increases with radius in \texttt{m12i} and decreases in \texttt{m12m} and \texttt{m12f}. We explore these differences further in Section~\ref{sec:where_form}.

The absolute number of giants with $z_{\rm form} > 5$ predicted within 10\,kpc of the solar neighborhood in the three simulations is (0.6, 2.3, and 1.2)$\,\times\,10^6$ for \texttt{m12i}, \texttt{m12f}, and \texttt{m12m}, respectively. For $z_{\rm form} > 10$, the corresponding numbers are (1.8, 7.0, and 6.3)\,$\times\,10^4$. We note that we have made no attempt to account for extinction, source confusion, or survey selection functions. Forthcoming mock catalogs for the three simulated galaxies studied here (Sanderson et al., in prep) will make it possible to do so.

\subsection{Metallicity}
\label{sec:metallicty}

\begin{figure*}
\includegraphics[width=\textwidth]{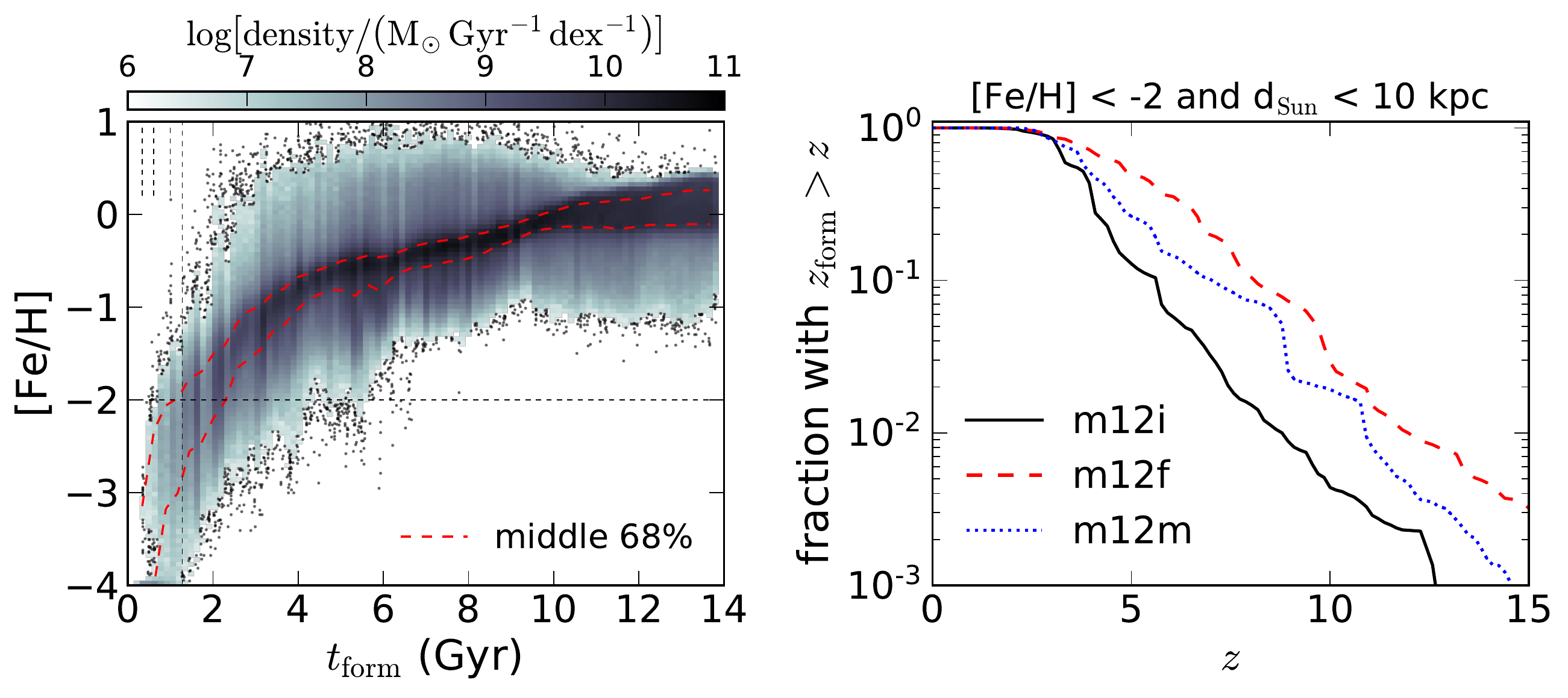}
\caption{{\bf Left}: Age - [$\rm Fe/H$] relation for all stars within the central 10 kpc of the m12i simulation. Vertical overdensities correspond to coherent bursts of star formation. Dashed vertical lines show $z=10,8,6,5$, from left to right. Stars forming at $z=5$ have a broad range of iron abundances, with $-4 < [\rm Fe/H] < -1.5$; stars with $\rm [Fe/H] = -2$ continue to form past $z=2$. {\bf Right}: Cumulative fraction of stars with $\rm [Fe/H] < -2$ and a distance $d < 10$\,kpc from the solar neighbourhood that formed before redshift $z$, for all three simulated galaxies in our sample. Depending on the details of the Milky Way's formation history, we predict that $(0.5-3)\%$ of metal-poor stars within 10\,kpc of the solar neighborhood formed before $z = 10$, and $(10-50)\%$ formed before $z = 5$.}
\label{fig:age_met}
\end{figure*}

Many ongoing searches for the oldest stars in the MW operate under the implicit assumption that the oldest stars are also the most metal-poor. We now investigate the relation between metallicity and age predicted by our simulations. 

Figure~\ref{fig:age_met} shows the age-[Fe/H] relation of stars that occupy the central 10\,kpc of the \texttt{m12i} simulation at $z=0$ (see \citealt{Lamberts_2018} for the age-metallicity relation for all metals). Qualitatively, most ancient stars are metal-poor, and most metal-poor stars are old. However, it is also clear that particularly for old stars, the relationship between age and $[\rm Fe/H]$ is not monotonic. For example, selecting stars with $\rm [Fe/H] = -2$ yields stars formed between $z\sim 8$ and $z\sim 1.5$, and stars with $z_{\rm form}\sim 5$ have $-3 \gtrsim \rm [Fe/H] \gtrsim -1$.

In the right panel of Figure~\ref{fig:age_met}, we show for all three simulations the fraction of metal-poor stars within 10\,kpc of the solar neighborhood that formed before a given redshift. For all three simulated galaxies, the majority of stars with $\rm [Fe/H] < -2$ formed before $z=2.5$; the redshift at which half of nearby metal-poor stars had formed is $z = 4-5$, while the redshift at which 10\% had formed is $z=6-8$. Thus, we predict that selecting stars with $\rm [Fe/H] < -2$ in a survey of the inner Galaxy will yield stars with median formation redshifts of $\sim$5, with of order 10\% forming before reionization.

The right panel of Figure~\ref{fig:age_met} also highlights the nontrivial effects of the details of MW-like galaxies' different early assembly histories on the age distributions of their oldest stars at $z=0$: the fraction of metal-poor stars that formed before a particular redshift varies by a factor of $\sim$6 between the earliest (\texttt{m12f}) and latest (\texttt{m12i}) forming galaxies. 

\begin{figure*}
\includegraphics[width=\textwidth]{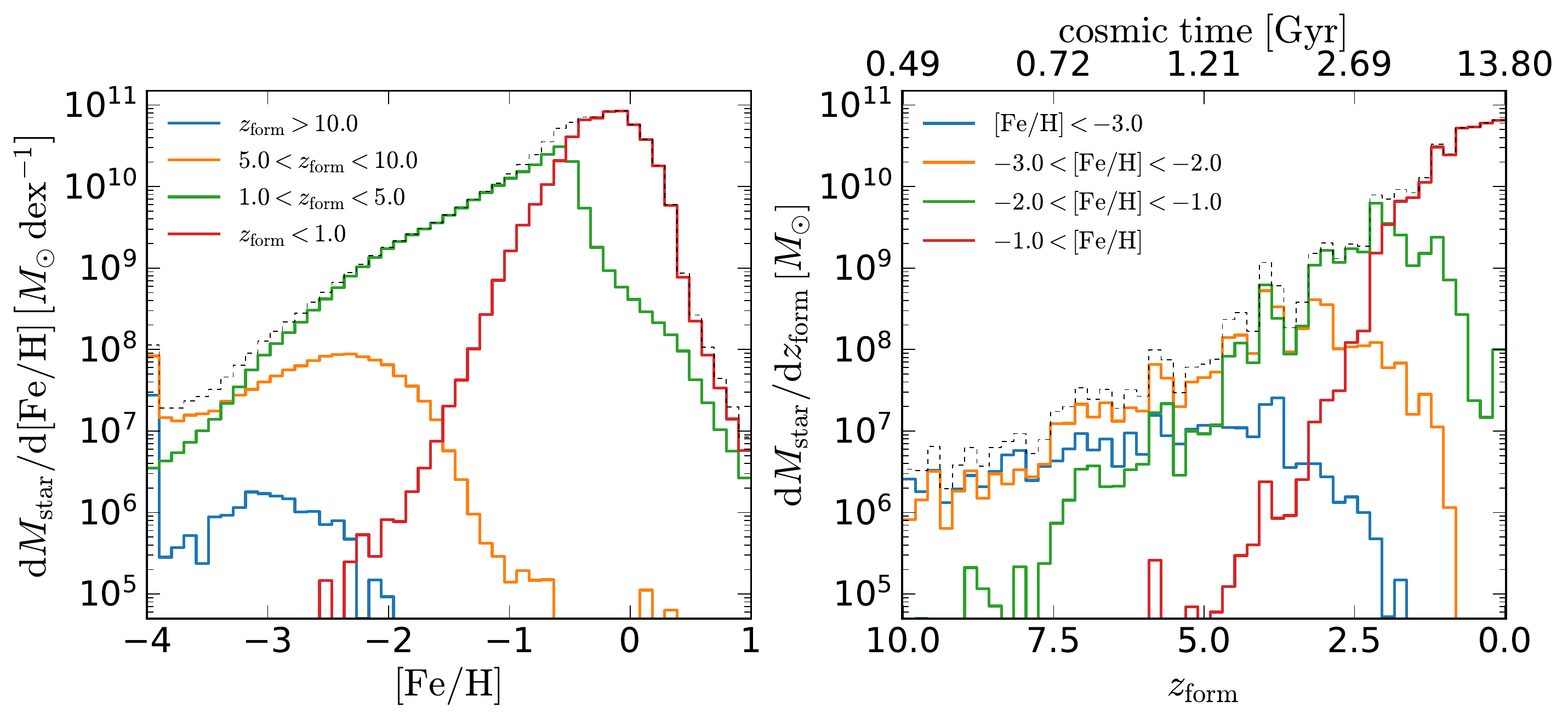}
\caption{{\bf Left:} Distribution of stellar metallicities and formation redshifts for stars in the central 10\,kpc of the simulation \texttt{m12i} at $z=0$. Colored histograms show individual metallicity/formation redshift bins; dashed black histogram shows total for all stars. Early-forming stars ($z_{\rm form} \gtrsim 5$) exhibit a broad range of metallicities peaked at $\rm -3 < [Fe/H] < -2$; they constitute $\sim$10\% of all stars with $\rm [Fe/H] < -2$ and almost half of all stars with $\rm [Fe/H] < -3$. {\bf Right:} stars with $\rm [Fe/H] > -3$ dominate over extremely metal poor stars at $z_{\rm form} \gtrsim 8$; for $z_{\rm form}=5$, there are a factor of $\sim$10 more stars with $\rm  [Fe/H]$ above $-3$ than below.}
\label{fig:feh_hist}
\end{figure*}

In Figure~\ref{fig:feh_hist}, we investigate the metallicity distributions of stars of different ages in more detail. The left panel shows the metallicity distribution for four different formation redshift intervals. Most stars formed before $z=10$ have $\rm [Fe/H] < -3$, while most stars with $5 < z_{\rm form} < 10$ have \ $\rm [Fe/H] < -2$. The most common metallicity for stars in the latter interval is $\rm [Fe/H] \sim -2.4$, but the distribution is quite broad. $\sim 20\%$ of stars with $z_{\rm form} > 5$ have $\rm [Fe/H] < -3$. The vast majority of stars formed at late time are enriched to higher metallicities: essentially all stars formed after $z=1$ have $[\rm Fe/H] > -1$. 

Figure ~\ref{fig:feh_hist} also shows that nearly 70\% of stars with $z_{\rm form} > 10$  (and 4\% of those with $5 < z_{\rm form} < 10$) have $[\rm Fe/H] = -4$; i.e., they formed from gas that had not been enriched above the metallicity floor at all. These metallicities clearly cannot be interpreted literally; fortunately, the majority of stars formed after $z\sim 8$ are enriched above this level. The latest-forming star particles that are in the central 10 kpc at $z=0$ and have pristine abundances form at $z = 4.3$ (\texttt{m12i}), $z=4.1$ (\texttt{m12f}) and $z=3.5$ (\texttt{m12m}).

The right panel shows the distribution of formation redshifts in bins of metallicity. Very low-metallicity stars with $\rm [Fe/H] < -3$ form over a broad range of redshifts, peaking at $z\sim 4$. There are roughly equal numbers of stars with metallicities above and below $\rm [Fe/H] = -3$ at $8 < z_{\rm form} <10$. The more enriched population begins to dominate at later redshifts, outnumbering the extremely metal-poor population by a factor of $\sim$10 at $z=5$ and $>$100 at $z=3$. Conversely, metal-rich stars are almost never ancient: only $\sim$0.1\% of stars formed before $z=5$ have $\rm [Fe/H] > -1$.

We do not show abundance distributions for elements other than iron, but we also find that all old stars that have been enriched above the metallicity floor exhibit enhanced abundance of $\alpha$ elements (e.g., Mg and Si) relative to solar values. Star particles formed before $z\sim 3$ typically have $\rm [\alpha/Fe]\sim 0.4$, while those forming after $z=1$ typically have $\rm [\alpha/Fe] \lesssim 0.25$. Enhancement in $\alpha$ elements is a necessary but not sufficient condition for a star particle to be ancient in our simulations: typical $[\rm \alpha/Fe]$ values do not begin to drop until $z\sim 3$ in all three simulations. Detailed abundance tracks for the simulated galaxies studied here will be presented in Wetzel et al., in prep.

\subsection{Kinematics}
\label{sec:kinematics}

\begin{figure*}
\includegraphics[width = \textwidth]{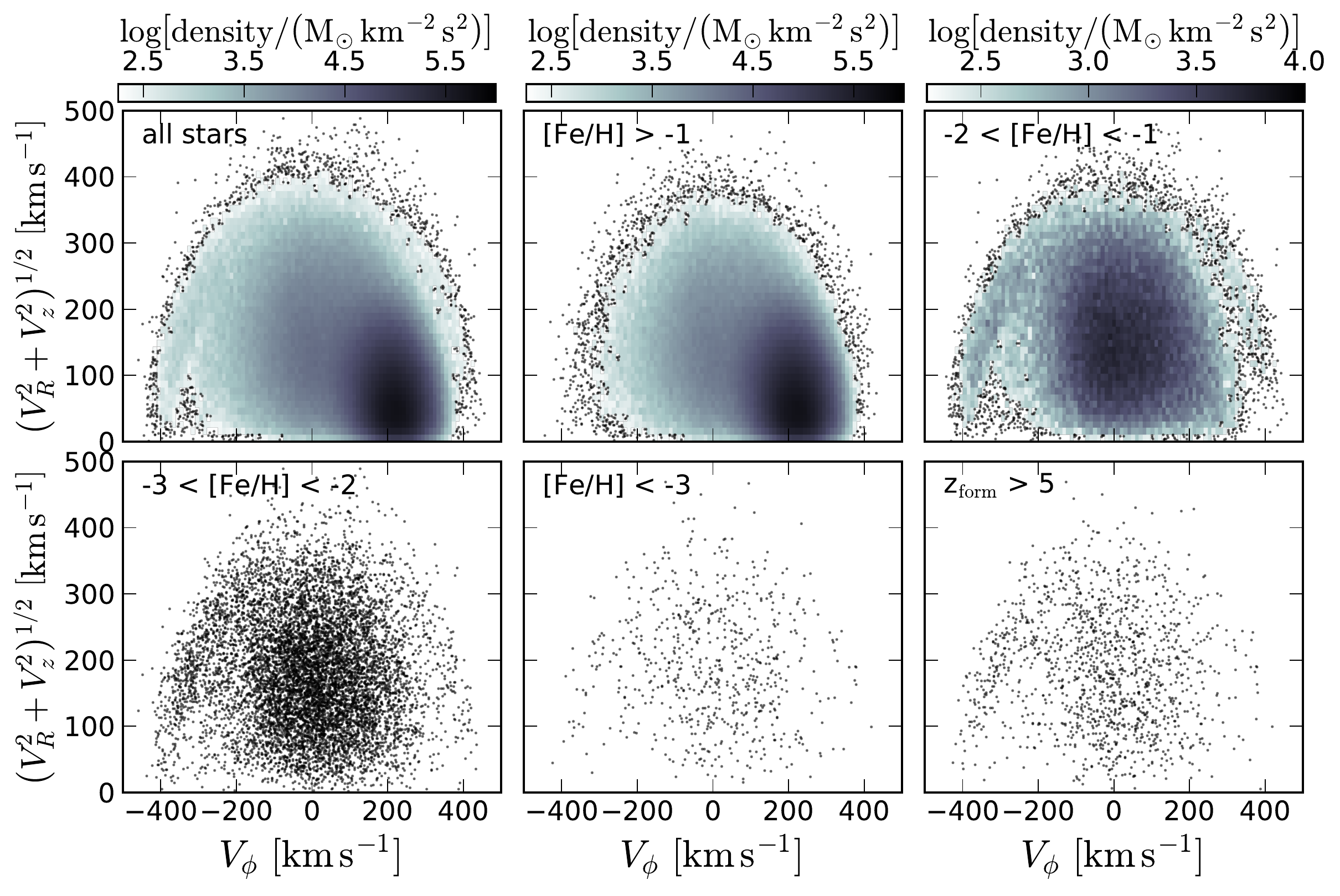}
\caption{Toomre diagrams of stars in the solar circle of the \texttt{m12i} simulation at $z=0$. We select stars in a cylindrical shell with $(6 < R/{\rm kpc} < 10)$ extending $\pm2$\,kpc from the disk midplane and plot all stars, stars in a range of decreasing metallicity bins, and stars that formed before $z=5$. In the top three panels, we plot density as a color scale when more than 5 star particles fall within a given pixel.
Most stars in the solar neighborhood are found on rotation-supported, disk-like orbits, but almost all of these stars have $\rm [Fe/H] > -1$. Metal-poor stars 
inhabit primarily dispersion-supported orbits. All old stars are dispersion-supported, with no significant coherent rotation. }
\label{fig:toomre}
\end{figure*}

We now examine the kinematics of old stars in our simulated galaxies and compare to the kinematics of the populations formed at later times. Figure~\ref{fig:toomre} shows Toomre diagrams for stars in \texttt{m12i}, separating all stars, stars in a range of decreasing metallicity bins, and the oldest stars. For consistency with work on the kinematics of stars in the MW solar neighborhood, we select stars in a cylindrical shell centered on $R=8$\,kpc; however, our results do not depend sensitively on the region in which stars are selected. The Toomre diagram compares the azimuthal rotation velocity of stars ($V_\phi$) with the speed along axes perpendicular to the main rotational motion. We define the cylindrical coordinate system such that the $+\hat{z}$ axis is aligned with the net stellar angular momentum vector. 

The upper left panel of Figure~\ref{fig:toomre} shows all stars. Most stars are in the disk, rotating coherently with mean rotation velocity $\left\langle V_{\phi}\right\rangle \sim 230$\,km\,s$^{-1}$. There is also a non-rotating, dispersion-supported population centered on $\left\langle V_{\phi}\right\rangle \sim 0$ that contains $<10$ percent of the total stellar mass for stars near the solar circle. Stars in this region of the Toomre diagram are typically referred to as having ``halo-like'' orbits. We refer to \citet{Bonaca_2017} for further discussion of stellar kinematics in the solar neighborhood of this simulation and comparison to observations of the MW. 

The upper middle panel shows only metal-rich stars, and the following three panels show stars in increasingly lower-$\rm [Fe/H]$ bins. Most stars with $\rm [Fe/H] > -1$ are in the rotation-supported disk population, while most stars that are more metal-poor are found on dispersion-supported orbits. This occurs because the disk in our simulated galaxies forms at $z\sim 1$ \citep{Ma_2017}\footnote{We note that the onset of disk formation is gradual, and its timing varies somewhat across our simulated galaxies \citep{GarrisonKimmel_2017}. We remain agnostic of the precise age of the MW's disk.}, when the mean iron abundance was $\rm [Fe/H]\sim-0.8$ (see Figure~\ref{fig:age_met}). Some substructure due to disrupted satellites can be seen in the Toomre diagrams for metal-poor stars, but there is little systematic difference between the kinematics of stars with $\rm -2 < [Fe/H] < -1$ and those with $\rm -3 < [Fe/H] < -2$ or $\rm [Fe/H] < -3$. Similarly, selecting only old stars ($z_{\rm form} > 5$; bottom right panel) yields dispersion-supported orbits indistinguishable from those of metal-poor stars formed at later times (before $z\sim 1$, after which most stars form in the disk). Having dispersion-supported kinematics is thus a necessary but not sufficient condition for a star in the simulated galaxies being ancient.

We find qualitatively similar results for the \texttt{m12f} and \texttt{m12m} simulations: the vast majority of the oldest stars are found on dispersion-supported orbits, but at fixed galactocentric radius, there is little difference between the kinematics of the oldest stars and stars formed at intermediate redshifts.

\section{Where did the oldest stars form?}
\label{sec:where_form}

\begin{figure*}
\includegraphics[width=\textwidth]{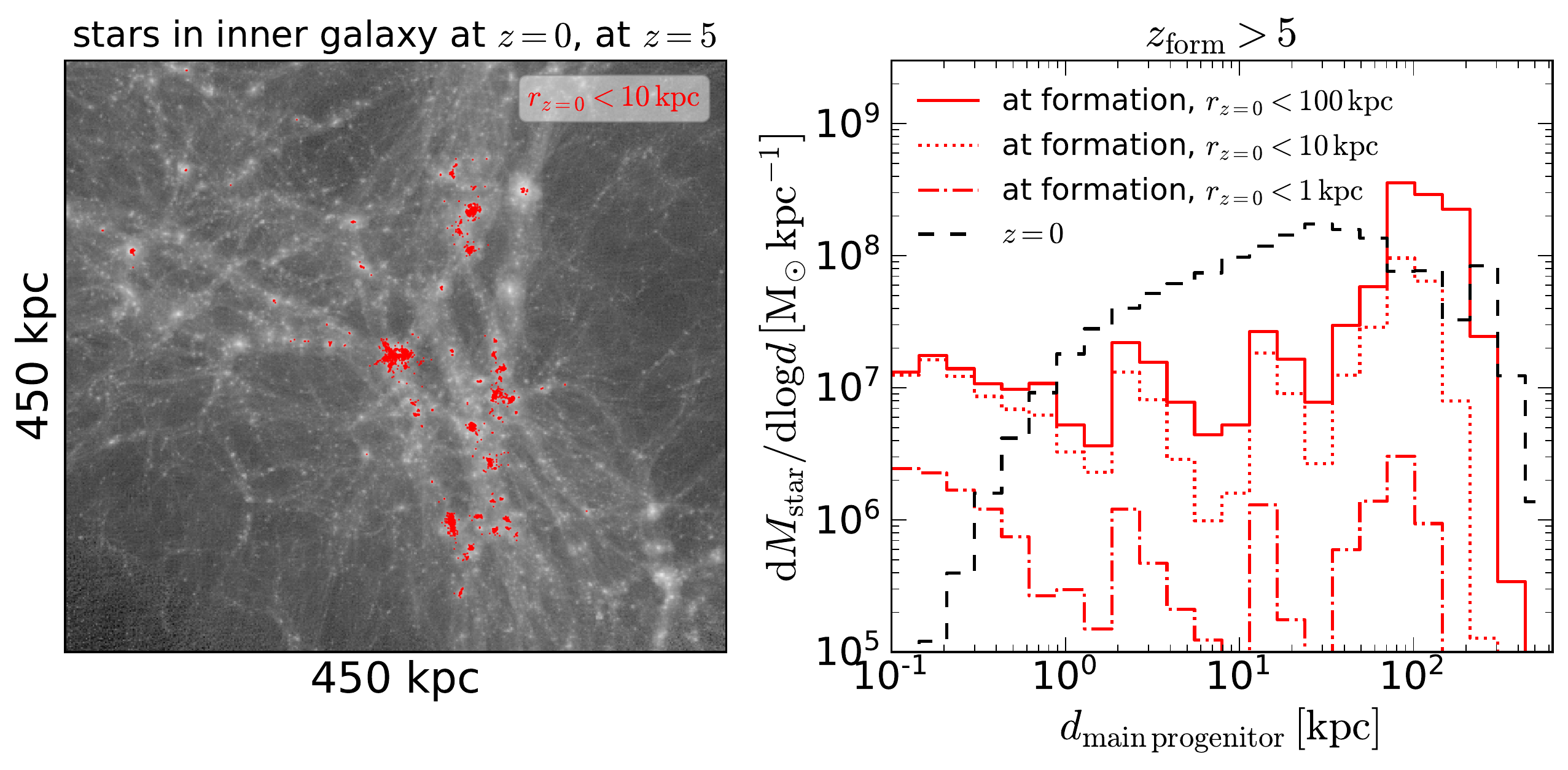}
\caption{\textbf{Left}: Red points show the oldest stars in the main galaxy in the \texttt{m12i} simulation -- i.e., stars within 10 kpc of the galactic center at $z=0$ -- traced back to $z=5$. Grayscale shows the projected dark matter surface density, integrated over a 400 kpc column. All units are physical, not comoving. At $z=5$, the main progenitor (center of left panel) contains less than half of the old stars that will end up in the inner galaxy by $z=0$. \textbf{Right}: Distances of the oldest stars from the main progenitor galaxy's center at the time of their formation (red) and at $z=0$ (black). The majority of the oldest stars in the galaxy at $z=0$ did not form there, but were deposited later through mergers. Stars in the outer halo at $z=0$ formed at even larger distances on average than those in the inner regions; the opposite is true for stars in the central 1 kpc.}
\label{fig:stars_at_z5}
\end{figure*}

We have shown that the FIRE simulations predict the oldest stars in MW-like galaxies to be less centrally-concentrated than later-forming stars, extending from the galactic center into the outer halo (Figure~\ref{fig:aitoff_m12i}, middle right panel). Previous works \citep[e.g.][]{ElBadry_2016, Angles_Alcazar_2017, Bonaca_2017, Sanderson_2017} have shown that mergers and secular dynamical processes, particularly outflow-driven fluctuations in the gravitational potential, can cause stars to migrate substantially after they form. We now investigate where the oldest stars formed, and how they arrived at their present-day spatial and kinematic distribution. 

\subsection{Defining formation sites}
To determine where the oldest stars in the simulated galaxies at $z=0$ formed, we tag star particles that formed before $z=5$ and are in the primary galaxy at $z=0$ (unless otherwise stated, within 10\,kpc of the galactic center) and then trace them back to the time of their formation. For each old star particle, we calculate the distance (always in physical units) from the main progenitor galaxy at the time of the star particle's formation, which we approximate as the first simulation output in which it appears.\footnote{Simulation outputs are saved every $\sim$20 Myr, so star particles are expected to migrate $\ll 1$\,kpc on average (with respect to their host galaxy at formation) between their formation and the first snapshot in which their positions are saved.} We determine the main progenitor in each snapshot as follows. 

At $z=0$, the main galaxy in the zoom-in region is identified as the galaxy with the highest stellar mass that is uncontaminated by lower-resolution dark matter particles. Its center is located using an iterative ``shrinking spheres'' method \citep{Power_2003}, wherein we recursively compute the center of mass of star particles in a spherical region, reducing the sphere's radius by $\sim$50\% and re-centering on the new stellar center-of-mass at each iteration. 

We then trace the main galaxy back through all simulation outputs. We define the main progenitor in terms of stellar mass, not halo mass.  In each output, we first calculate the ``expected'' location of the galaxy at that time by extrapolating backwards from its position and velocity in the next snapshot. We then repeat the shrinking-spheres centering algorithm, beginning with a sphere of radius 20\,kpc centered on the expected location. We define the galaxy identified in this way as the ``main progenitor'' in each output.

This procedure ensures that the same galaxy is followed consistently through all snapshots. At high redshift, it is often the case that the main progenitor is not the most massive galaxy in the zoom-in region, since different galaxies grow at different rates. To ensure that the method produces the desired behaviour, we also verify that for every simulation output, more than 50\% of the star particles in the main progenitor were also in the main progenitor in the previous output. We note that due to scatter in the $M_{\rm star}-M_{\rm halo}$ relation, the main progenitor identified in this way is not necessarily the same as the main progenitor identified from the dark matter merger tree \citep[e.g.][]{Fitts_2018}.\footnote{We find that defining the main progenitor using star particles generally yields the same results as when dark matter particles are used at low redshift, but not at $z\gtrsim 3$. At very high redshifts ($z \gtrsim 10$), it can even occur that the main progenitor defined from the dark matter merger tree does not contain any stars.}  

\subsection{In situ vs. ex situ formation}
In Figure~\ref{fig:stars_at_z5}, we trace the oldest stars in the \texttt{m12i} simulation back to their locations at high redshift. The left panel shows the spatial distribution at $z=5$ of star particles that inhabit the central 10 kpc of the galaxy at $z=0$. It is clear that at $z=5$, these stars were \textit{not} part of a single coherent population, but were distributed throughout of order 100 distinct lower-mass galaxies that subsequently merged. In \texttt{m12i}, the main progenitor (shown in the center of the left panel of Figure~\ref{fig:stars_at_z5}) \textit{was} already the most massive galaxy in the zoom-in region, but it still contained less than half of the old stars that would make their way into the central galaxy by $z=0$. We note that many of the dark matter halos in the left panel of  Figure~\ref{fig:stars_at_z5} that do not contain red points \textit{do} host stars; they simply never merge with the main progenitor, and are swept up in the Hubble flow by $z=0$. 

The right panel of Figure~\ref{fig:stars_at_z5} shows the distance from the main progenitor of stars that formed before $z=5$ at the time of their formation (i.e., some time before $z=5$; red) and at $z=0$ (black). We plot distributions of formation distance for stars within different concentric shells at $z=0$. The dotted red histogram shows that, even considering only stars in the central 10\,kpc at $z=0$, a majority of old stars formed at distances of order 100 kpc from the main progenitor. When stars in the outer halo at $z=0$ are included (solid red line), only a few percent of the oldest stars formed in the main progenitor. The most distant-forming stars in the central 10\,kpc at $z=0$ formed at distances of $\sim$250\,kpc, while some stars in the outer halo formed at distances in excess of 400\,kpc from the main progenitor. 

Comparing the black and red histograms in the right panel of Figure~\ref{fig:stars_at_z5}, it is also evident that old stars that formed in the central regions of the main progenitor have on average moved to larger distance by $z=0$. We investigate this in more detail in Section~\ref{sec:migration}.

\begin{figure}
\includegraphics[width=\columnwidth]{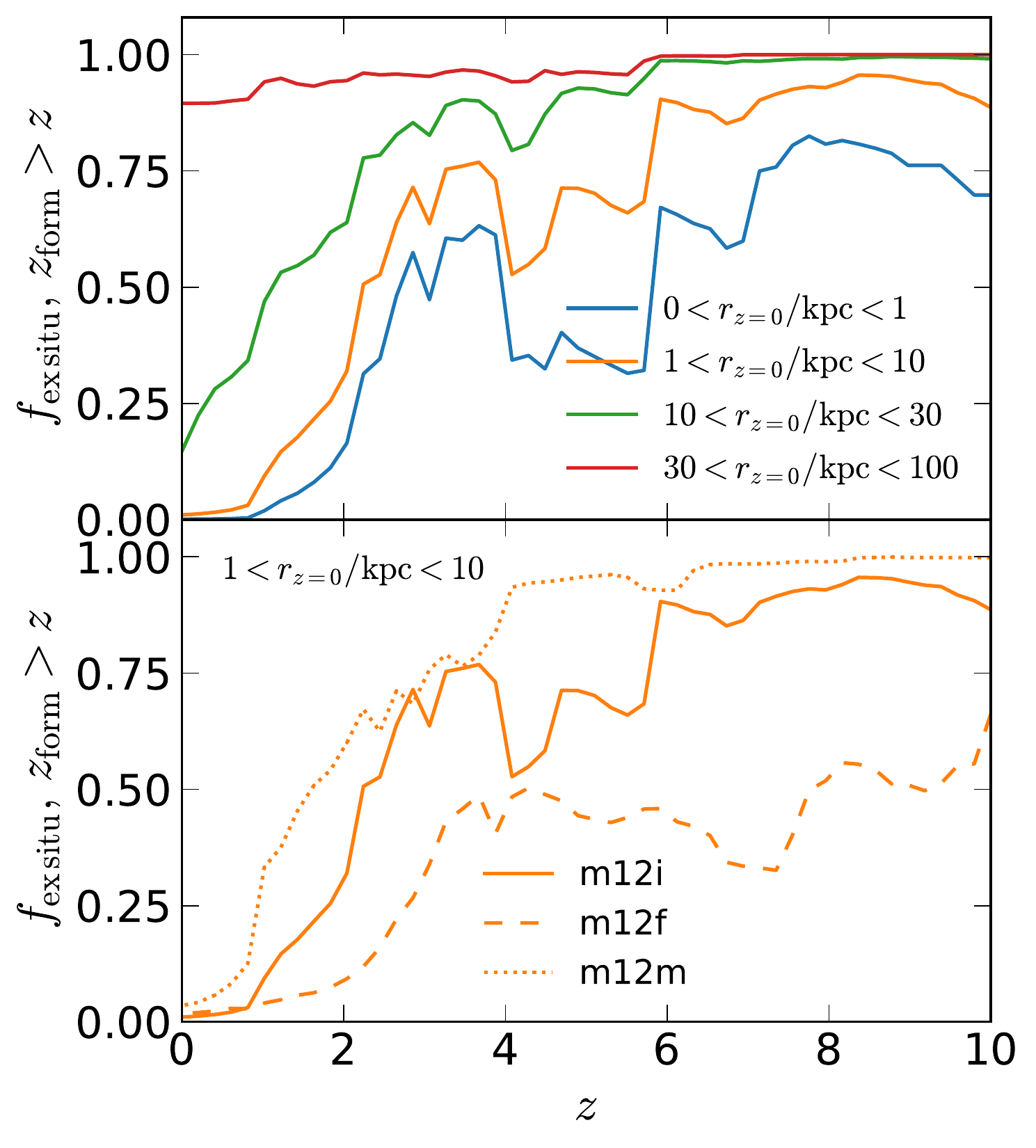}
\caption{\textbf{Top}: Cumulative fraction of stars forming before redshift $z$ that formed {\it ex situ} (i.e., more than 20 physical kpc from the main progenitor) in \texttt{m12i}. We plot stars in different radial bins at $z=0$ separately. The fraction of stars formed {\it ex situ} increases with $r_{z=0}$ for stars of all ages. Most stars with $z_{\rm form} \gtrsim 3$ -- even those in the inner galaxy at $z=0$ -- formed {\it ex situ}. \textbf{Bottom}: Same, but for all three simulated galaxies, considering a single radial bin. {\it Ex situ} stars generally dominate the population formed before $z\sim 3$}, but there is significant variation in the detailed formation histories of the three simulations.
\label{fig:f_ex_situ}
\end{figure}

Figure~\ref{fig:f_ex_situ} shows explicitly how the fraction of stars formed {\it ex situ} (which we define as forming more than 20\,kpc from the main progenitor) varies with formation time. In the top panel, for \texttt{m12i} only, we show the {\it ex situ} fraction in different radial bins. At all formation redshifts, the {\it ex situ} fraction increases monotonically with radius: more centrally-concentrated stars were preferentially formed {\it in situ}. However, even for the central bin ($r_{z=0}<10$\,kpc), in \texttt{m12i} it is only after $z=2$ that stars formed {\it in situ} begin to dominate over those accreted at later times. Intriguingly, this panel also shows that a nonzero fraction of stars in the outer halo ($30 < r_{z=0}/{\rm kpc} < 100$), including some formed at late times, formed {\it in situ}. We find that these stars primarily form from already-outflowing, feedback-driven gas clouds, similar to results found in other simulations \citep{Purcell_2010, Cooper_2015, Elias_2018}. The origin and fate of these stars will be explored in more detail by Yu et al. (in prep).

The bottom panel of Figure~\ref{fig:f_ex_situ} compares the fraction of stars within 10\,kpc at $z=0$ that formed {\it ex situ} in all simulations. For all three simulated galaxies, at least half of the stars born before $z\sim 4$ were formed {\it ex situ} and subsequently accreted. However, the fraction of stars of a given age formed {\it ex situ} varies substantially across the three simulations: in \texttt{m12m}, {\it ex situ} stars dominate at $z_{\rm form}>2$, while in \texttt{m12f}, of order half of all stars formed {\it in situ} up to $z = 10$. 

\begin{figure}
\includegraphics[width=\columnwidth]{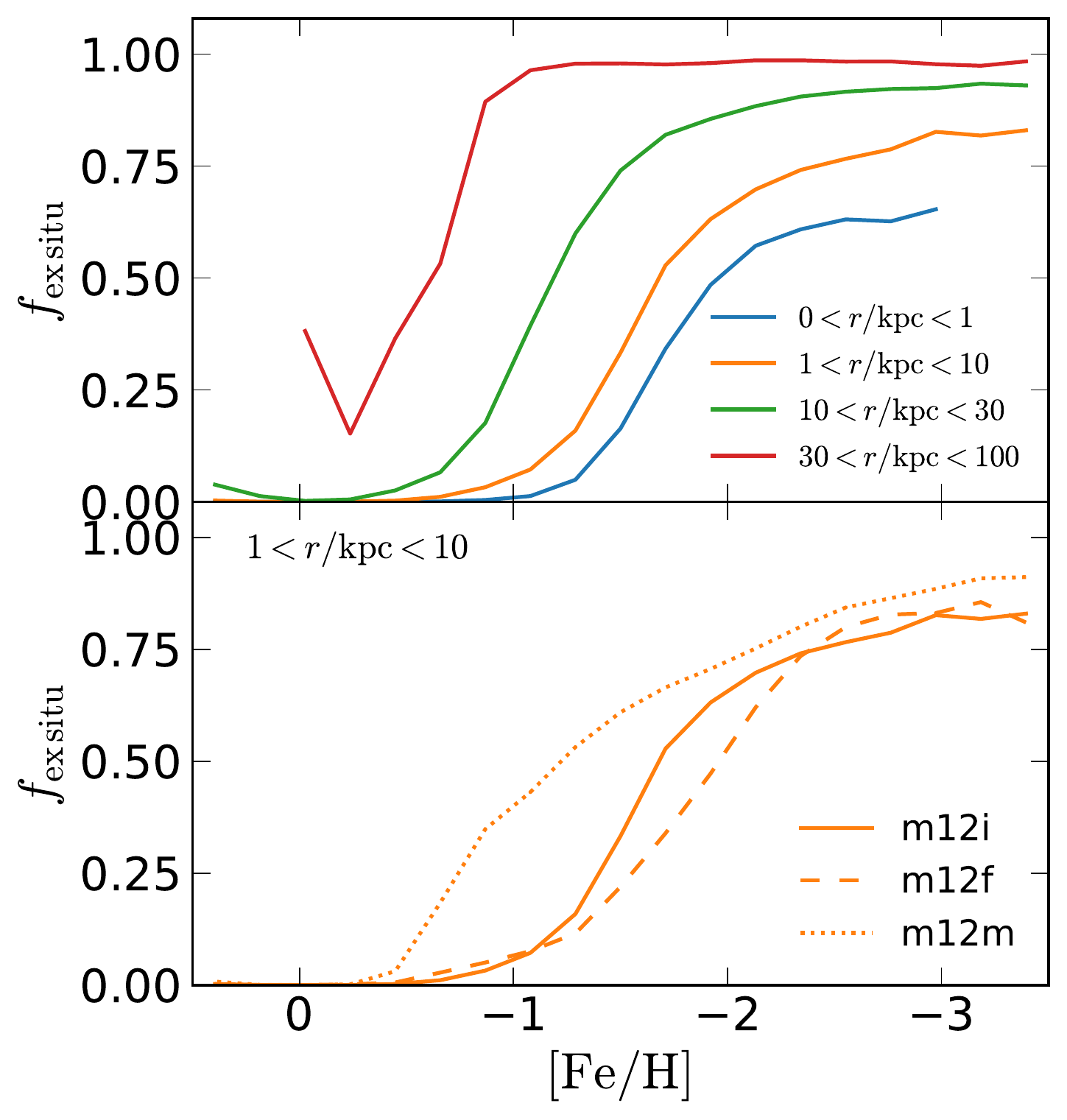}
\caption{\textbf{Top}: Fraction of stars with a given metallicity that formed {\it ex situ} and were subsequently accreted in \texttt{m12i}. We plot stars in different radial bins at $z=0$ separately. The fraction of stars formed {\it ex situ} increases with $r$ for stars of all metallicities. Most stars with $[\rm Fe/H] \lesssim -1.5$ -- even those in the inner galaxy -- formed {\it ex situ}. \textbf{Bottom}: Same, but for all three simulated galaxies, considering a single radial bin. {\it Ex situ} stars dominate the low-metallicity population in all simulations.}
\label{fig:f_ex_situ_feh}
\end{figure}

Figure~\ref{fig:f_ex_situ_feh} shows the fraction of stars formed {\it ex situ} as a function of metallicity rather than formation redshift. Even within the solar circle, most metal-poor stars formed {\it ex situ} and were subsequently accreted. The metallicity below which the majority of stars in the inner galaxy formed {\it ex situ} varies between -1.9 in \texttt{m12f} and -1.2 in \texttt{m12m}. This suggests that simply selecting metal-poor stars -- at any radius -- will predominantly yield stars that formed {\it ex situ}.

\begin{figure}
\includegraphics[width=\columnwidth]{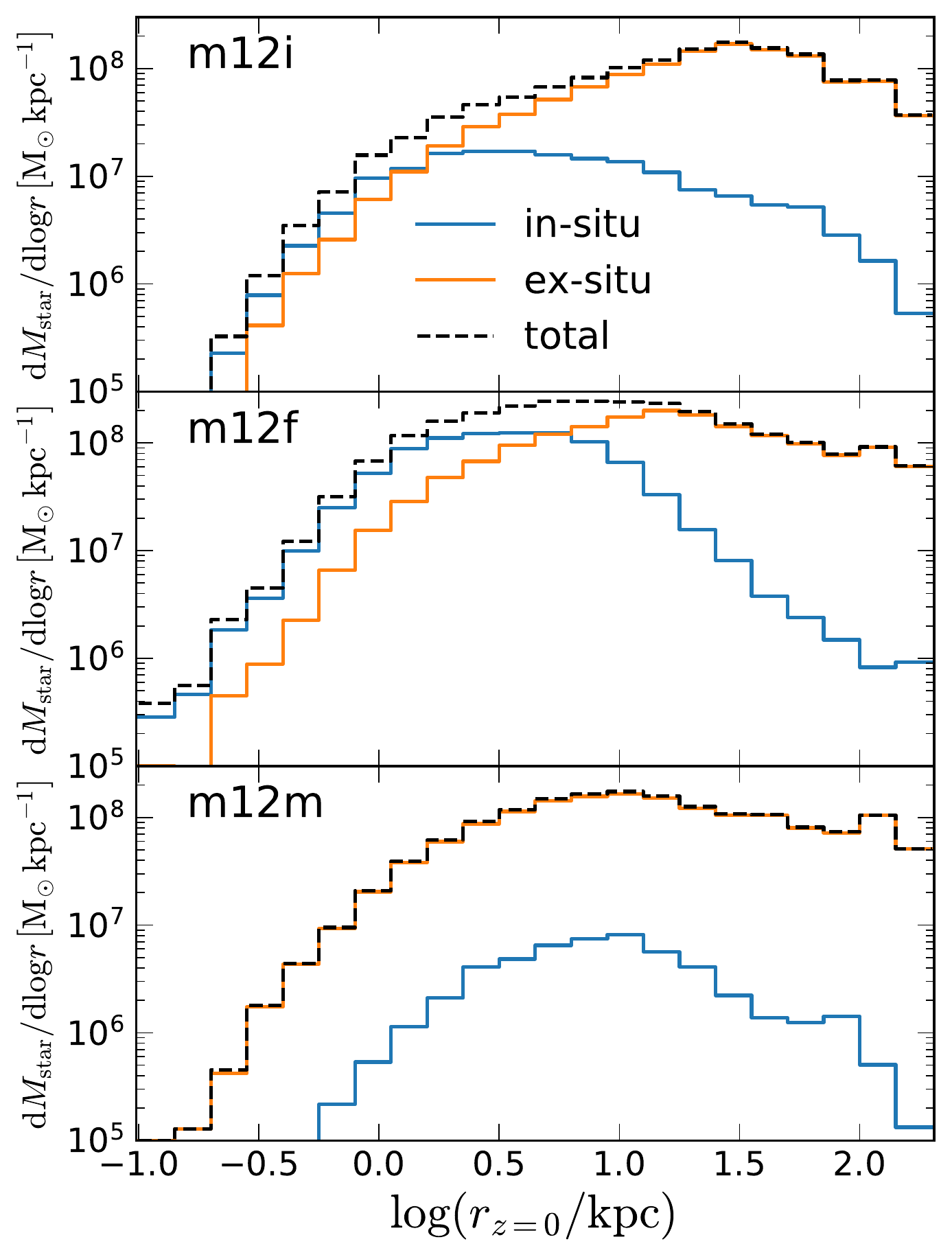}
\caption{Radial distribution of old stars ($z_{\rm form} > 5$) at $z=0$ for all three simulated MW-like galaxies. Stars formed {\it in situ} are generally more centrally concentrated than those formed {\it ex situ}; they dominate over ex-situ stars in the central $\sim$1\,kpc (\texttt{m12i}) and $\sim$5\,kpc (\texttt{m12f}). In \texttt{m12m}, the main progenitor contained only a small fraction of the total stellar mass at $z > 5$, so {\it ex situ} stars dominate at all radii.}
\label{fig:ex_situ_hists}
\end{figure}

In Figure~\ref{fig:ex_situ_hists}, we show the spatial distribution at $z=0$ of the oldest stars in all three simulations, separating stars that formed {\it in situ} and those that formed {\it ex situ}. Old stars formed {\it in situ} are generally more centrally concentrated: in \texttt{m12i}, they outnumber the {\it ex situ} stars in the central $\sim$1\,kpc, but {\it ex situ} stars become dominant at $r>5$\,kpc. The situation is qualitatively similar in \texttt{m12f}, which formed earlier, but {\it in situ} stars make up a larger fraction of the old population and thus dominate out to $\sim$5\,kpc. Finally, in \texttt{m12m}, the {\it ex situ} population is completely dominant at all radii. 

In \texttt{m12m}, the main progenitor does not contribute the oldest coherent population of old stars in the galaxy, as several other halos host galaxies more massive than the main progenitor at $z\sim 5$ that eventually merge with the main progenitor. In this sense, ``main progenitor'' is not an especially meaningful title at high redshift: although the main progenitor can always be uniquely defined, there is no guarantee that it is the \textit{dominant} progenitor. Similarly, the distinction between {\it in situ} and {\it ex situ} stars ceases to be physically meaningful before a main progenitor is established. 

\begin{figure*}
\includegraphics[width=\textwidth]{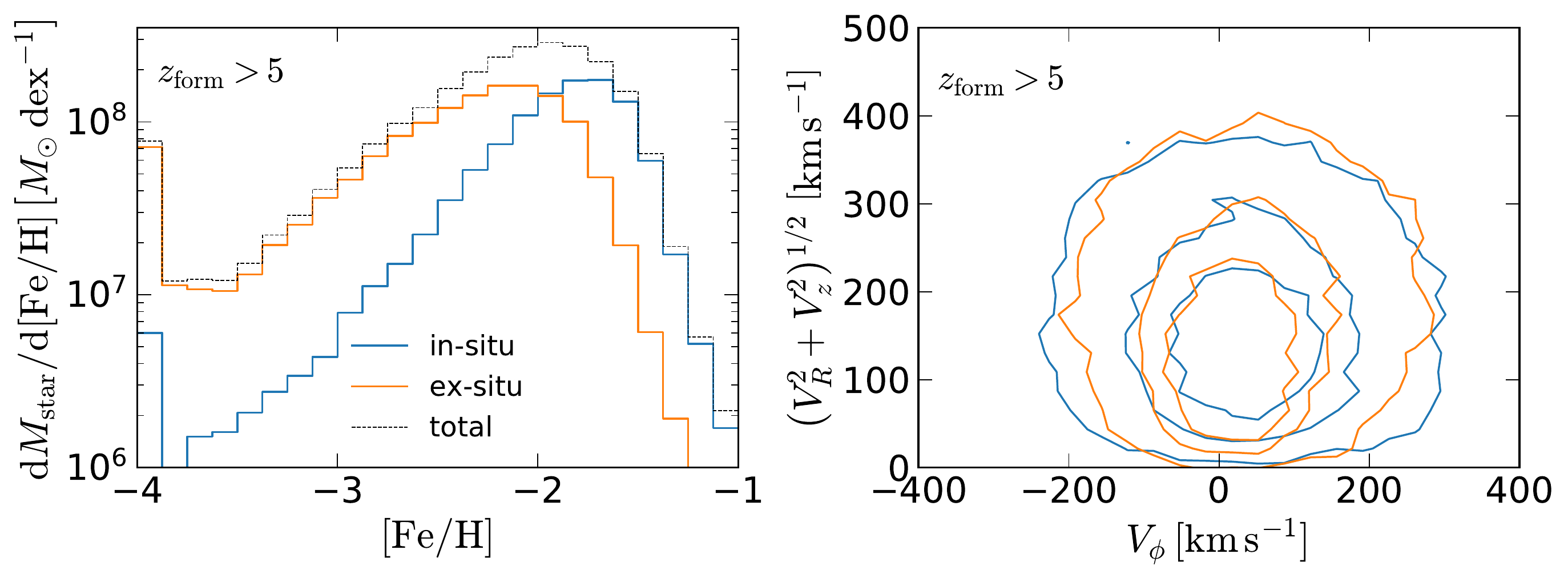}
\caption{Metallicty distribution (left) and Toomre diagram (right) of old stars in the \texttt{m12f} simulation, which has the largest old {\it in situ} population of the simulations we study (Figure~\ref{fig:ex_situ_hists}). We divide stars that are in the main galaxy ($r<30$\,kpc) at $z=0$ into {\it in} vs. {\it ex situ} populations with a distance cut of $d_{\rm main\,progenitor}=20$\,kpc at the time of their formation. Old stars that formed {\it in situ} have higher metallicities on average ($\left\langle {\rm \left[Fe/H\right]}\right\rangle =-1.7$) than those that formed {\it ex situ} ($\left\langle {\rm \left[Fe/H\right]}\right\rangle =-2.2$), but the two populations have identical dispersion-supported kinematics.}
\label{fig:in_vs_ex_situ_feh}
\end{figure*}

In Figure~\ref{fig:in_vs_ex_situ_feh}, we consider the possibility of distinguishing between the {\it in situ} and {\it ex situ} populations based on $z=0$ observables. Here we show the simulation \texttt{m12f}, which has a coherent main progenitor up to the highest redshifts and thus offers the best chance of exhibiting a coherent {\it in situ} population at late times. Consistent with previous work \citep{Johnston_2008, Corlies_2013}, we find that the mean metallicity of old stars is higher for the  old {\it in situ} population, reflecting the fact that the galaxy mass-metallicity relation remains steep at high redshift \citep{Ma_2016}. It is thus in principle possible to assign a known old star a probability of having formed {\it in situ} based on its metallicity; for example, an old star in \texttt{m12f} with $[\rm Fe/H]=-1.5$ has a $>90$\% probability of having formed {\it in situ}. In practice, separating 
the {\it in situ} and {\it ex situ} populations based on $\rm [Fe/H]$ alone is entirely infeasible, both due to the difficulty of reliably age-dating such moderately low-metallicty stars and because the metallicity distributions of the MW's old {\it in situ} and {\it ex situ} stars are not known a priori. We discuss possible avenues for distinguishing these populations based on abundances of other elements in Section~\ref{sec:discussion}.

In the right panel of Figure~\ref{fig:in_vs_ex_situ_feh}, we compare Toomre diagrams of the old {\it in situ} and {\it ex situ} populations in \texttt{m12f}. Contours are separated by factors of two and show the velocity-space surface density of old stars, normalized by the total mass of each population. This figure shows that the {\it in situ} and {\it ex situ} populations have essentially  indistinguishable kinematics: both are dispersion supported and retain no obvious memory of their formation. This is perhaps unsurprising, since violent relaxation during hierarchical merging at later times generously redistributes energy between stellar orbits. \citet{Bonaca_2017} found similar results for stars in the solar neighborhood of \texttt{m12i}, including those formed and accreted at later times. Similarly, \citet{Sanderson_2017} found {\it ex situ} and {\it in situ} stars in these simulated galaxies to be spatially co-located and well-mixed at $z=0$, such that standard attempts to separate the populations using photometric profiles perform poorly. 

Our simulations predict that a large fraction of the oldest stars in the Milky Way -- including those near the center today -- did not form locally, but were accreted at later times after forming in external subhalos. This in part explains why our model predicts a reduced fraction of old stars towards the Galactic center (Figure~\ref{fig:aitoff_m12i}) and overwhelming dispersion-supported kinematics for old stars (Figure~\ref{fig:toomre}): while stars that form at later times $(z \lesssim 2)$ are born from low-orbital energy gas near the center of the primary halo's gravitational potential, stars deposited in mergers have higher energies and are spread over a large range of radii \citep[e.g.][]{Bullock_2005, Sharma_2017}. 

But as we discuss below, this is only part of the story.

\subsection{Outward migration of old stars}
\label{sec:migration}

\begin{figure*}
\includegraphics[width=\textwidth]{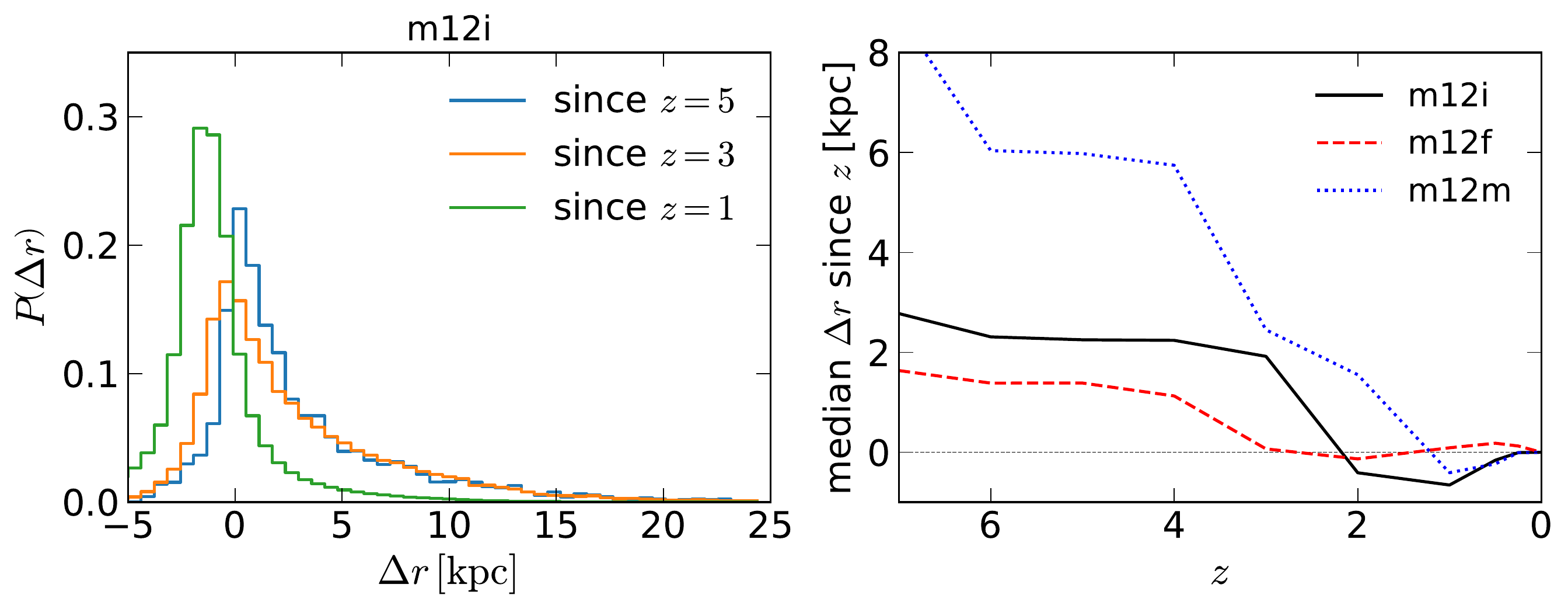}
\caption{{\bf Left}: Distribution at $z=0$ of distances stars in the \texttt{m12i} simulation have migrated since $z=5$, $z=3$, and $z=1$. At each redshift, we tag all star particles within 5\,kpc of the galactic centre and then trace them to $z=0$, computing $\Delta r = r_{z=0} - r_{z\,{\rm initial}}$ (so positive $\Delta r$ represents outward migration). Fluctuations in the shallow gravitational potential at high redshift can add energy to stellar orbits, causing older stars to migrate to larger radii. These fluctuations have largely died down by $z=1$, so later-forming stars remain centrally concentrated; in fact, contraction of the central potential during the formation of the bulge causes stars to migrate {\it inward} at later times. {\bf Right}: median $\Delta r$ as a function of time for all three simulations. Outward migration ceases at $z\sim 2-1$ and is weaker for simulations dominated by a single progenitor (see Figure~\ref{fig:f_ex_situ}). }
\label{fig:migration}
\end{figure*}

It is evident from the right panel of Figure~\ref{fig:stars_at_z5} that stars move both inward and outward between their formation and $z=0$. As we discuss in the previous section, inward motion is a necessary result of hierarchical assembly through mergers. We now investigate in more detail what drives stars to move outward after they form; to this end, we focus on stars formed {\it in situ} near the galactic center. 

To assess how old stars migrate as a function of redshift, we tag star particles that are near the galactic center ($r<5$\,kpc) at some high redshift $z_{\rm initial}$, follow them to $z=0$, and measure $\Delta r = r_{z=0} - r_{z = z_{\rm initial}}$ (so outward migration  corresponds to positive $\Delta r$). Because stellar orbits are not necessarily circular, some nonzero $\Delta r$ is expected simply due to random changes in orbital phase from one snapshot to the next. To minimize the scatter due to this effect, we always calculate both $r_{z=0}$ and $r_{z = z_{\rm initial}}$ in five simulations snapshots spread over $\sim$100 Myr; for each particle, we use the median radius over these 5 snapshots. 

We show values of $\Delta r$ for $z_{\rm initial}=5,3,$ and 1 in Figure~\ref{fig:migration}. The left panel shows distributions of $\Delta r$ for star particles in \texttt{m12i}. Most stars have moved outward since $z=5$, with a median $\Delta r=2.5$\,kpc and $\sim$25\% of stars migrating outward more than 6\,kpc. Since $z=3$, most stars have migrated outward, but typical $\Delta r$ values are slightly lower than since $z=5$. Finally, most stars within $r<5$\,kpc migrate {\it inward} between $z=1$ and $z=0$. The right panel of Figure~\ref{fig:migration} shows the median $\Delta r$ since redshift $z$ for all three simulations. In all cases stars formed {\it in situ} before $z\sim 4$ migrate outward on average by $z=0$, but the details differ nontrivially across different simulations. 

\begin{figure}
\includegraphics[width=\columnwidth]{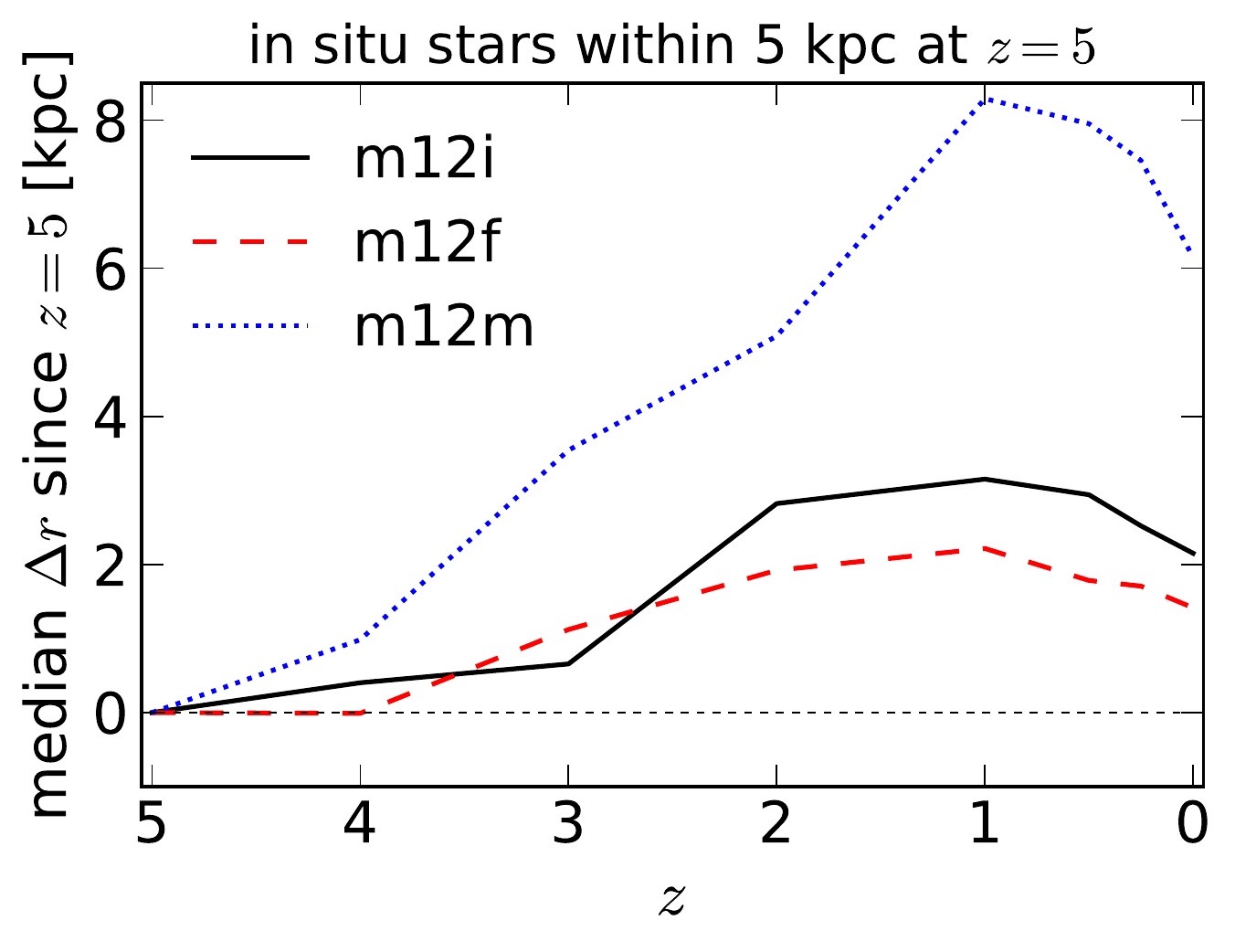}
\caption{At $z=5$, we tag star particles with $r < 5\,$kpc that formed {\it in situ}. We then follow the same star particles until $z=0$, plotting their median outward migration. In all simulations, stars migrate outward until $z\sim 1$, and then partially migrate back inward as the potential contracts at $z \lesssim 1$. }
\label{fig:migration_since_z5}
\end{figure}

Whereas Figure~\ref{fig:migration} shows how much stars that were near the galactic center at a given redshift have migrated by $z=0$, Figure~\ref{fig:migration_since_z5} follows the migration of one particular set of star particles -- those that formed {\it in situ} and are near the galactic center at $z=5$ -- as a function of time. For all three simulations, old {\it in situ} stars (on average) migrate outward continually between $z=5$ and $z=1$, and then migrate inward between $z=1$ and $z=0$. 

Previous works \citep[e.g.][]{Read_2005, Stinson_2009, Maxwell_2012, ElBadry_2016, ElBadry_2017} have shown that in low-mass galaxies, a time-varying gravitational potential, generated primarily by stellar-feedback driven gas outflows, can drive stars formed {\it in situ} into the stellar halo. This process has also been shown to create cores in low-mass galaxies' dark matter density profiles \citep{DiCintio_2014, Onorbe_2015, Chan_2015}. It is not generally thought to operate in $L_\star$ galaxies, as it relies on rapid outflows launched by bursty star formation to drive impulsive changes in the depth of the gravitational potential \citep{Pontzen_2012, Pontzen_2014}. Star formation is typically not bursty at late times in MW-mass galaxies, but it \textit{is} at higher redshift \citep[e.g.][]{Sparre_2017, FG_2018}. Indeed, the shallow potentials of MW-progenitors at early times render them susceptible to the same type of feedback-driven net outward stellar migration found by \citet{ElBadry_2016} to operate in low-mass galaxies at late times.

Stars cease migrating outward after $z\sim 2-1$ (right panel of Figure~\ref{fig:migration} and Figure~\ref{fig:migration_since_z5}) because fluctuations in the gravitational potential have largely died down by this time: mass accumulation in the galactic center due to continued gas inflow deepens the potential until feedback-driven outflows can no longer impulsively evacuate dynamically significant quantities of gas from the galactic center. The stronger (weaker) outward migration of old stars in \texttt{m12m} (\texttt{m12f}) can also be understood as a consequence of the simulation's less (more) dominant main progenitor at high redshift: it is easier to drive large-scale potential fluctuations, due both to outflows and to mergers, in a shallow potential. 

After $z\sim 1$, contraction of the potential due to continued accumulation of baryons at high density \citep[e.g.][]{Blumenthal_1986} actually drives stars to migrate \textit{inward},  partially undoing outward migration at earlier times. This inward migration is somewhat weaker in \texttt{m12f}, perhaps because this galaxy forms a weaker bulge than the other two simulations, so that the central potential contracts less at later times \citep[see][]{GarrisonKimmel_2017}. 

\citet{Chan_2015} studied the evolution of the central dark matter density profile in FIRE simulation of MW-like galaxies and found similar redshift evolution to what we find for the oldest stars (e.g. compare Figure~\ref{fig:migration_since_z5} to their Figure 5): stellar feedback-driven potential fluctuations remove dark matter from a galaxy's inner regions until $z\sim 1$, but contraction of the halo at later times due to continued accretion of baryons in the absence of bursty star formation partially undoes this effect by $z=0$. 

\section{Comparison to previous work}
\label{sec:sim_compare}

A number of previous studies \citep{Scannapieco_2006, Salvadori_2007, Salvadori_2010, Tumlinson_2010, Ishiyama_2016, Griffen_2018} have used dark matter only simulations combined with semi-analytic prescriptions for the formation sites of the first stars to model the distribution of old stars in MW-like galaxies at $z=0$. In qualitative agreement with this work, these studies have predicted old stars to be concentrated near the Galactic Center. We emphasize, however, that because the population of old stars in the inner galaxy is diluted at $z=0$ by later-forming stars (by a factor of $\sim$10,000 to 1; see Figure~\ref{fig:aitoff_m12i}) it may be more efficient to search for old stars at higher galactic latitudes. We also note that most studies based on dark matter only simulations do not account for destruction of substructure due to the strong tidal field of the baryon overdensity in the inner halo \citep[e.g.][]{GarrisonKimmel_2017a} and also cannot model baryon-driven fluctuations in the potential that drive stars outward, so we expect the oldest stars to be somewhat more dispersed than predicted by these works. 

Our study most closely resembles the recent work of \citet{Starkenburg_2017}, who studied the $z=0$ distribution of  the oldest stars in the APOSTLE simulations \citep{Sawala_2016} of paired halos selected to resemble the Local Group. These authors found that  50 percent of the most metal-poor stars, which they defined as $\rm [Fe/H] < -2.5$, form before $z = 5.3$. We find similar results: for our three halos, 50 percent stars with $\rm [Fe/H] < -2.5$ form before $z = 4.1$ (\texttt{m12i}), $z = 5.0$ (\texttt{m12f}) and $z = 4.8$ (\texttt{m12m}). They find that 90 percent of the most metal-poor stars form before $z = 2.8$; we find corresponding redshifts $z = 2.8$ (\texttt{m12i}), $z = 2.9$ (\texttt{m12f}) and $z = 3.0$ (\texttt{m12m}). 

We also find reasonable agreement with Starkenburg et al. when we reproduce their Figure 2, which shows the fraction of the oldest and most metal-poor stars compared to the total population as a function of galactic radius (see our Figure~\ref{fig:aitoff_m12i} for a rough comparison); our three halos all fall within the range of ancient fractions spanned by their simulations. Both APOSTLE and FIRE find that the fraction of the most metal-poor stars that are old declines with radius. 

However, the fraction of all stars that are old in the central few kpc is lower than the median value found by \citet{Starkenburg_2017} for all three of our simulated galaxies, by a factor of $\sim$3-4 on average. We also find the fraction of all stars that are ancient to increase by roughly an order of magnitude between the galactic center and the solar neighborhood, while they find it to be nearly flat. In other words, old stars in our simulations are less centrally concentrated than in APOSTLE, at least within $r \lesssim 10$\,kpc. Similarly, we find that the fraction of metal-poor stars in the inner galaxy is lower by a factor of $\sim$2 in our simulations; this disagreement is somewhat weaker than that for ancient stars, likely because a significant fraction of the metal-poor stars are formed {\it ex situ}. 

We suspect that old stars in our simulations are less centrally concentrated due to energetic feedback processes that drive outward the stars formed near the center of the high-redshift MW progenitor's shallow potential. The APOSTLE simulations do not attempt to model cold ($T\ll 10^4$\,K) gas and allow star formation to occur at relatively low densities ($n_{\rm H}\sim 10^{-1}\,{\rm cm^{-3}}$), a factor of 10,000 lower than the star formation density threshold of $n_{\rm H}=10^3\,{\rm cm^{-3}}$ adopted in the FIRE model. This causes star formation in their simulations to be less spatially and temporally clustered than in models that attempt to resolve the cold ISM, preventing the formation of cores in the dark matter density profiles of their low-mass galaxies \citep{Oman_2015}. Because the mechanism we propose for driving old stars formed {\it in situ} to larger radii relies in part on the same feedback-driven potential fluctuations that lead to the creation of these cores, we expect it to be less efficient in simulations that do not form cores. 

\section{Summary and Discussion}
\label{sec:discussion}

\begin{figure*}
\includegraphics[width=\textwidth]{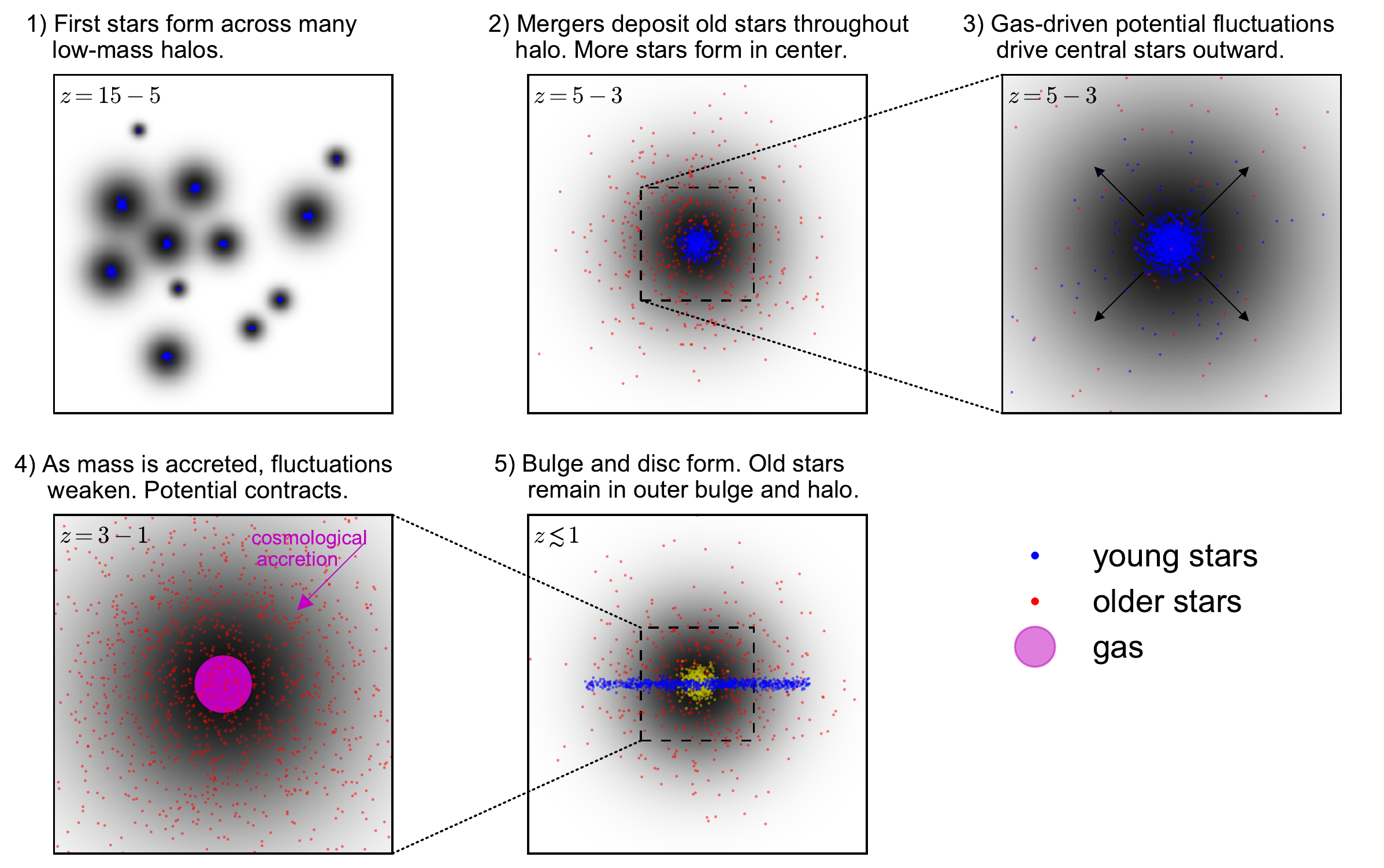}
\caption{Schematic illustration of the Galaxy's early assembly. The earliest-forming stars are much less centrally-concentrated at late times than stars in the disk and bulge because (a) many are formed {\it ex situ} and are deposited in the outer proto-halo in mergers, and (b) the oldest stars formed {\it in situ} are driven to larger radii by outflow-driven fluctuations of the shallow gravitational potential at $z\gtrsim 3$, which couple energy to the orbits of collisionless particles, as in dwarf galaxies.}
\label{fig:schematic}
\end{figure*}

We have studied the spatial distribution, kinematics, metallicity, and formation history of the oldest stars in FIRE cosmological baryonic zoom-in simulations of three MW-mass disk galaxies. Figure~\ref{fig:schematic} provides a schematic picture of the processes that lead to the observed distribution of the oldest stars at $z=0$. Our main results are as follows:

\begin{enumerate}
\item \textit{Spatial distribution of old stars at $z=0$}: The absolute number density of ancient stars ($z_{\rm form} > 5$) is highest near the galactic center, in the bulge and inner stellar halo. However, because stars that form at later times ($z\lesssim 2$) are {\it more} centrally concentrated at $z=0$, the fraction of all stars along a given line of sight that are ancient is in fact \textit{lowest} for sight lines towards the inner galaxy (Figure ~\ref{fig:aitoff_m12i}). The fraction of metal-poor stars that are ancient generally decreases with galactocentric radius, but with significant scatter between different simulations. 

\item \textit{Kinematics of old stars}: The oldest stars in our simulated galaxies are all on dispersion-supported, halo-like orbits, exhibiting no rotational support (Figure~\ref{fig:toomre}). The same is true for most metal-poor stars and most stars formed before $z\sim 1$, so dispersion-supported kinematics are a necessary but insufficient condition for a star to be ancient. Among old stars, stars formed {\it in situ} (i.e., within the main progenitor) and stars formed {\it ex situ} (in an external galaxy that was subsequently accreted) have indistinguishable kinematics (Figure~\ref{fig:in_vs_ex_situ_feh}).

\item \textit{Rapid enrichment}: The first generation of stars formed in the simulations rapidly enrich the gas surrounding their formation sites to $[\rm Fe/H]\sim -3$, so that a majority of ancient stars have metallicities of $-3 < [\rm Fe/H] < -2$ (Figures~\ref{fig:age_met} and~\ref{fig:feh_hist}) and enhanced [$\alpha/$Fe]. Because these moderately low metallicty stars are more abundant than stars with extremely low metallicity ($[\rm Fe/H] < -3$), they represent promising targets for surveys attempting to identify the oldest stars in the Galaxy, {\it if} their ages can be reliably measured. Conversely, the most metal-poor stars are not uniformly old: stars with $[\rm Fe/H] < -3$ continue to form until $z\sim 3$. 

\item \textit{In situ vs. ex situ formation}: Our simulations predict that a dominant fraction of the ancient stars in the MW at $z=0$ -- even those near the Galactic center -- did not form {\it in situ}, but formed in external galaxies, often more than 100 kpc from the main progenitor, and were deposited in the halo through hierarchical mergers at $5 \lesssim z \lesssim 1$ (Figure~\ref{fig:stars_at_z5}). Even in the central 10\,kpc, these {\it ex situ} stars dominate the old, metal-poor population: most stars with $z_{\rm form} \gtrsim 2$ formed {\it ex situ}. The same is true for most stars with $[\rm Fe/H] \lesssim -1.5$ (Figure~\ref{fig:f_ex_situ} and~\ref{fig:f_ex_situ_feh}). Accreted stars do not penetrate as deep into the potential as stars formed {\it in situ} at late times (Figures~\ref{fig:stars_at_z5} and ~\ref{fig:ex_situ_hists}). 
This is one of the two primary reasons old stars are less centrally concentrated at $z=0$ than stars formed at later times. 

\item \textit{Outward migration after formation}: Even old stars that form {\it in situ} and are initially centrally concentrated migrate outward by $z=0$. This outward migration is qualitatively similar to the processes proposed to create dark matter cores and drive stars in low-mass galaxies outward even at later times \citep{ElBadry_2016}: fluctuations in the gravitational potential, driven by large-scale gas inflows/outflows as well as mergers, add energy to stellar orbits. These fluctuations die out by $z\sim 2$, so that stars formed at later times do not migrate outward much, and in fact migrate inward as the potential contracts at later times (Figures~\ref{fig:migration} and~\ref{fig:migration_since_z5}).
\end{enumerate}

\subsection{Prospects for identifying the oldest stars}
\label{sec:future}

Our simulations predict that a substantial fraction of stars with $\rm [Fe/H]\lesssim -2$ formed before $z=5$ and are now spread throughout the Galaxy. Magnitude-limited surveys like RAVE, GALAH, APOGEE, LAMOST, and SEGUE find metal-poor stars in abundance. For example, RAVE detected 480,000 stars with SNR $>$ 20 at intermediate spectral resolution ($R\approx 9000$; \citealt{Kunder_2017}), about 6000 of which have $-4 \leq \rm [Fe/H]\leq -2$. We expect LAMOST to deliver an even larger set of very metal poor stars, albeit at lower spectroscopic resolution ($R\approx 2000$; \citealt{Li_2015}). GALAH has observed 550,000 stars to date at high resolution ($R\approx 30,000$; \citealt{DeSilva_2015}) for which 500 stars with [Fe/H] $<$ -2 have SNR $>$ 50 required for useful measurements of heavy metal abundances. Comparable numbers of metal-poor stars have been observed at high resolution and SNR by APOGEE \citep{Majewski_2017, FernandezAlvar_2017}. {\it Gaia}-RVS is expected to increase the known population of very metal-poor stars by more than an order of magnitude \citep{Robin_2012}.

Photometric surveys have also proved effective for identifying metal-poor stars. Although spectroscopic follow-up is required to measure detailed abundances and kinematics, photometric surveys can reliably identify metal-poor stars more efficiently and at larger distances than spectroscopic surveys, and they are sensitive to dwarfs as well as giants. The fact that metal-poor stars are bluer has long been used to photometrically identify candidate ancient stars \citep[e.g.][]{Wallerstein_1962}, and even broadband photometry, (e.g., from SDSS; \citealt{Ivezic_2008}) can yield metallicities accurate at the 0.1 dex level for metal-rich stars ($[\rm Fe/H] \gtrsim -1$). For moderately metal-poor stars ($[\rm Fe/H]\gtrsim -2.5$), intermediate-band Stromgren photometry can reliably measure metallicities with a precision of a few tenth of a dex \citep{Stromgren_1963, Arnadottir_2010}. Stars with still lower metallicities can be efficiently identified both with narrow-band photometry targeting the Calcium H \& K lines \citep{Beers_1985, Christlieb_2002, Hill_2017, Starkenburg_2017b, Youakim_2017} or mid-infrared photometry targeting broad absorption bands \citep[e.g.][]{Schlaufman_2014, Casey_2015}. State-of-the-art surveys searching for extremely metal-poor stars can photometrically identify stars with $[\rm Fe/H]< -3$ with a success rate of up to $\sim$25\%; the majority of their higher-metallicity contaminants still have $[\rm Fe/H]< -2$ \citep{Youakim_2017}. Finally, LSST is predicted to find tens of thousands of RR Lyrae stars throughout the Local Group \citep{Oluseyi_2012}; these span a wide range of metallicities but are all relatively old ($\gtrsim 10$\,Gyr). 

Compared to the search for ancient stars at the low metallicity limit, less effort has been expended to find and study those at $-3 \lesssim \rm [Fe/H] \lesssim -2$. This owes primarily to the difficulty of identifying the oldest members of this population. {\it If} higher-metallicity ancient stars can be reliably identified, targeting them offers some advantages over UMP stars. As we have demonstrated (e.g. Figure~\ref{fig:feh_hist}), ancient stars with  $-3 \lesssim \rm [Fe/H] \lesssim -2$ are predicted to be much more numerous than UMP stars of the same age. At fixed apparent magnitude, more elemental abundances can be measured for stars with higher metallicity, providing a more complete probe of the yields of their progenitor stars and, in exceptional cases, enabling age-dating with more precise chemical clocks. On the other hand, the gas clouds from which more metal-rich ancient stars formed were likely enriched by a larger number of supernovae, so the interpretation of these stars' abundance patterns in terms of first-star yields is more challenging. 

The most significant challenge facing efforts to use ancient stars for Galactic archaelogy stems from the difficulty of measuring accurate stellar ages; i.e, identifying which stars are actually ancient. At present, there is {\it no} reliable method for age-dating individual old stars with better than $\sim$10\% accuracy (see \citealt{Soderblom_2010}, for a review); this means, for example, that stars formed at $z=10$ cannot be reliably distinguished from stars formed at $z=4$, at least not without appealing to galactic chemical enrichment models. Even for globular clusters, where information in many parts of the HR diagram can be leveraged to simultaneously constrain the age of the stellar population, state-of-the-art uncertainties on absolute stellar ages are of order $\pm 1$\,Gyr (\citealt{Chaboyer_2017}, or see \citealt{Choksi_2018}, for a compilation). 

Investments in 3D stellar atmospheric models and measurement of many heavy element abundances \citep[e.g.][]{Nissen_2017, Chiavassa_2018} may somewhat improve age constraints on stars for which exceptionally high-resolution and high-SNR data are available. Measuring accurate stellar ages will nevertheless present a significant challenge for the foreseeable future. This is due in large part to persistent and fundamental uncertainties in stellar interior models, such as shortcomings in mixing-length theory \citep{Joyce_2017} and non-negligible uncertainties in the experimentally-measured opacities used for stellar interior calculations \citep[e.g.][]{Bailey_2015}. 

Ongoing and planned asteroseismic missions are predicted to deliver mass measurements of some RGB stars \citep{Chaplin_2014, Rauer_2014, Ricker_2015} accurate enough to, in principle, yield ages that are accurate at the $5-10$\% level. However, uncertainties in mass loss rates on the RGB \citep[e.g.][]{Miglio_2012} are currently large enough to prevent measurement of initial stellar masses with better than $\sim$10\% accuracy, and asteroseismic mass estimates for most RGB stars carry similar uncertainties \citep{Stello_2015}. Typical uncertainties on asteroseismic ages thus remain of order 15\% \citep{Silva_Aguirre_2015}, comparable to the uncertainties in age estimates of giants from the spectroscopic [C/N] ratio \citep[e.g.][]{Ness_2016} and age estimates for main-sequence stars with white dwarf companions \citep{Fouesneau_2018}. The {\it precision} of such ages is likely to improve with the influx of better data, but the uncertainties in stellar models discussed above currently set a more uncompromising limit on the accuracy of absolute ages, at the 7-10\% level. We also note that asteroseismic mass determinations for low-metallicity stars may be subject to some biases \citep[e.g.][]{Epstein_2014}, and precise [C/N]-based ages cannot be obtained at low metallicity. 

In lieu of direct age estimates from stellar physics, a possible alternate route for identifying ancient stars at intermediate metallicity is to appeal to specific metal-producing populations (e.g. $r$-process in core collapse SNe) that are likely to have occurred before other high-yield populations (e.g. $s$-process in AGBs, Fe peak elements in SNe Ia). While the precise origin of the $r$-process continues to be controversial \citep{Winteler_2012, Arcones_2013}, with recent work suggesting a prompt channel in neutron star mergers \citep{Belczynski_2017}, most models suggest that $r$-process enhancement with respect to Fe peak and $s$-process elements provides strong evidence a star is ancient \citep{Frebel_2015}.

Are there stars at $\rm [Fe/H]\sim -2$ that have enhanced $r$-process elements compared to $s$-process elements? The literature is sparse, but some candidates do exist. A newly revealed example is the retrograde halo star RAVE J153830.9-180424 with $\rm [Fe/H]=-2.09$, enhanced $\rm [\alpha/Fe]=0.34$, enhanced $\rm [Eu/Fe]=1.27$ and suppressed $\rm [Ba/Fe]$ \citep{Sakari_2018}. Detailed modeling of the actinide elements (Th, U) with respect to the other heavy elements supports the idea of this star being ancient, albeit with systematic uncertainties of several Gyr. Similar abundance patterns have also been found for some stars in metal-poor globular clusters \citep[GCs;][]{Sneden_1997, Worley_2013, Roederer_2015}. 

Indeed, although their formation remains poorly understood theoretically, GC age constraints purely from stellar models are consistent with a large fraction of metal-poor GCs forming before $z=5$ \citep{Chaboyer_1995, Krauss_2003, Bastian_2017}. Because GCs are too small to be properly resolved in the simulations we study here, we cannot yet predict their ages and $z=0$ spatial distribution a priori. Dynamical friction arguments \citep[e.g.][]{Capuzzo_1993, Carlberg_2017} suggest that since-disrupted GCs should be found in the inner bulge, and recent observation appear to support this idea \citep{Minniti_2016, Schiavon_2017}. Modeling the self-consistent formation and evolution of GCs in a cosmological context is a promising avenue for future work. 

Given the nature and extent of existing stellar surveys and the difficulty of reliably age-dating old stars, it is unsurprising that metal-enriched ancient stars have yet to be identified in significant numbers. If such stars {\it can} be reliably identified in the future, then their spatial distribution will serve as a constraint on feedback models: because these stars were formed when the MW's progenitor's shallow potential was more susceptible to feedback-driven outflows than at late times, the dynamical imprint of baryon-driven potential fluctuations in our simulations is much larger for ancient stars than for stars formed at late times. 

\section{Acknowledgements}
We have benefited from fruitful discussions with K.C. Freeman,  A. Karakas, S. Loebman, C.F. McKee, and S. Tabin. 
KE acknowledges support from a Berkeley graduate fellowship, a Hellman award for graduate study, and an NSF Graduate Research Fellowship. 
JBH acknowledges a Miller Professorship from the Miller Institute, UC Berkeley. JBH is also supported by an ARC Laureate Fellowship from the Australian Government.
AW was supported by NASA through grants HST-GO-14734 and HST-AR-15057 from STScI.
EQ and KE are supported by a Simons Investigator Award from the Simons Foundation and by NSF grant AST-1715070.
DRW is supported by a fellowship provided by the Alfred P. Sloan Foundation.
MBK acknowledges support from NSF grant AST-1517226 and CAREER grant AST-1752913 and from NASA grants NNX17AG29G and HST-AR-13888, HST-AR-13896, HST-AR-14282, HST-AR-14554, HST-AR-15006, HST-GO-12914, and HST-GO-14191 from STScI. 
Support for PFH was provided by an Alfred P. Sloan Research Fellowship, NASA ATP Grant NNX14AH35G, and NSF Collaborative Research Grant \#1411920 and CAREER grant \#1455342.
CAFG was supported by NSF through grants AST-1412836, AST-1517491,AST-1715216, and CAREER award AST-1652522, by NASA through grant NNX15AB22G, and by a Cottrell Scholar Award from the Research Corporation for Science Advancement.
DK was supported by NSF grants AST-1412153 and AST-1715101 and the Cottrell Scholar Award from the Research Corporation for Science Advancement.
Support for SGK was provided by NASA through Einstein Postdoctoral Fellowship grant number PF5-160136 awarded by the Chandra X-ray Center, which is operated by the Smithsonian Astrophysical Observatory for NASA under contract NAS8-03060.
We ran numerical calculations on the Caltech compute cluster ``Wheeler,'' allocations TG-AST130039 \& TG-AST150080 granted by the Extreme Science and Engineering Discovery Environment (XSEDE) supported by the NSF, and the NASA HEC Program through the NAS Division at Ames Research Center and the NCCS at Goddard Space Flight Center.
The analysis in this paper relied on the python packages \texttt{NumPy} \citep{vanderwalt_2011}, \texttt{Matplotlib} \citep{Hunter_2007}, and \texttt{AstroPy} \citep{Astropy_2013}.




\bibliographystyle{mnras}

\appendix
\section{Effect of diffusion coefficient and resolution}
\label{sec:diff_coeff}

\begin{figure}
\includegraphics[width=\columnwidth]{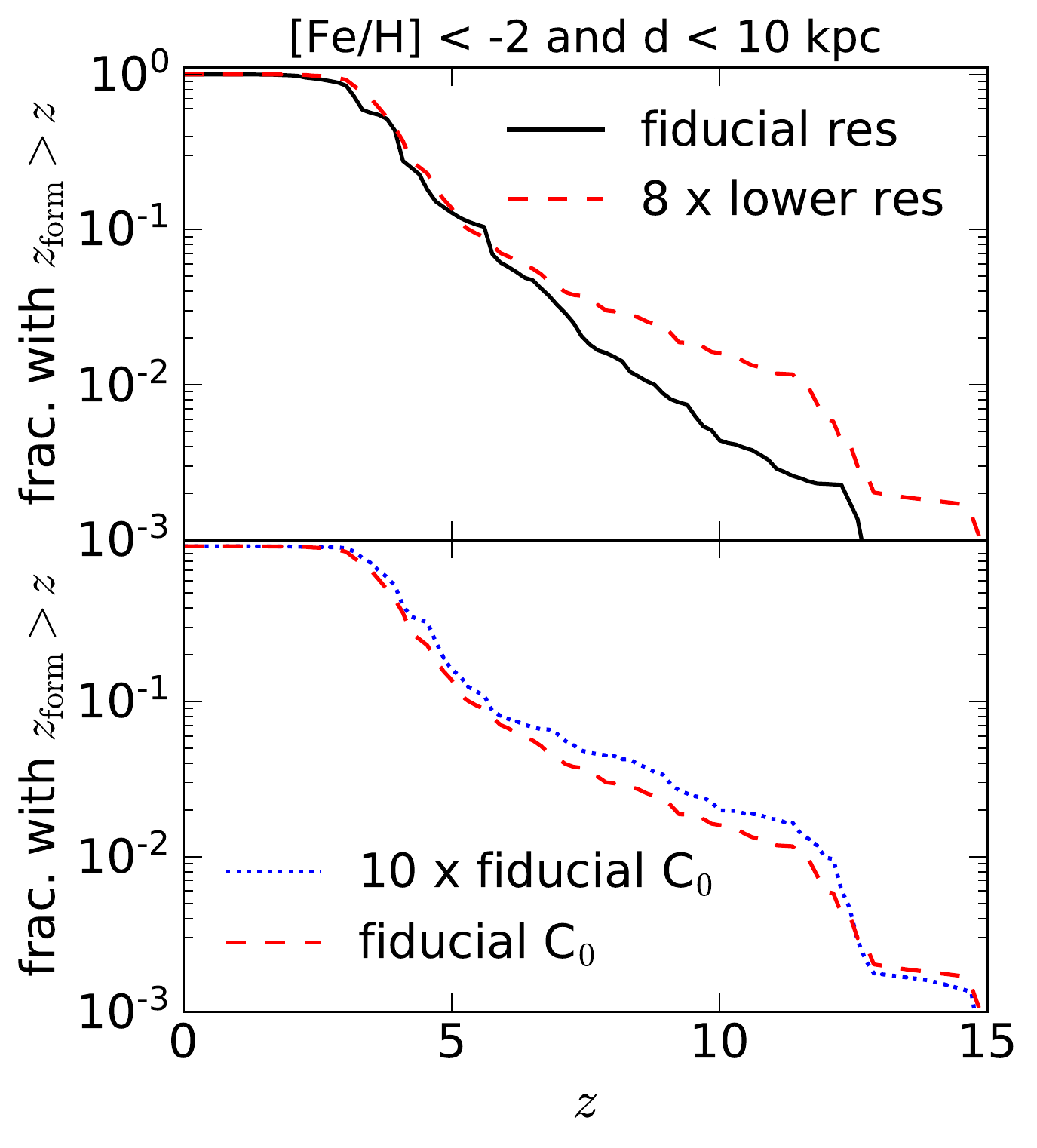}
\caption{Effect of varying the mass resolution (top) and diffusion coefficient (bottom) on the fraction of metal-poor stars that formed before a particular redshift. Black line shows the default \texttt{m12i} simulation, identical to the black line in the right panel of Figure~\ref{fig:age_met}. Red dashed line shows the same simulation run with 8 times lower mass resolution, but the same diffusion coefficient. Blue dotted line shows a run with 8 times lower mass resolution than the default \texttt{m12i} and 10 times higher diffusion coefficient.}
\label{fig:res}
\end{figure}

Figure~\ref{fig:res} shows the effect of varying the simulation mass resolution (top panel) and the turbulent diffusion coefficient (bottom panel). The quantity plotted is the fraction of stars within 10\,kpc of the solar neighborhood with $\rm [Fe/H] < -2$ that formed before a given redshift, identical to the right panel of Figure~\ref{fig:age_met}. We compare three different versions of the \texttt{m12i} simulation. The fiducial version analyzed throughout the text (black line) has a baryon mass resolution of $m_{\rm b} = 7070 M_{\odot}$ and uses a diffusion coefficient C$_{0}\approx 0.003$ (see \citealt{Hopkins_2017} for details). 

In the top panel, we compare this run to one with 8 times lower mass resolution ($m_{\rm b}=56600 M_{\odot}$) and the same diffusion coefficient and critical density for star formation as the fiducial run. For stars formed after $z\sim 7$, the fraction of metal-poor stars that formed before a given redshift agrees between the two resolution levels within $\sim$20\%. At higher $z_{\rm form}$, the predictions of the two resolution levels disagree more significantly, with a maximum discrepancy of a factor of 5. The fraction of metal-poor stars that are very old is higher in the lower-resolution run. We find that the total age distribution of ancient stars is similar in the high- and low-resolution runs. Typical metallicities of the most ancient stars are higher in the high-resolution run, indicating that gas is enriched somewhat more rapidly at high resolution.

The bottom panel compares two runs with the same mass resolution ($m_{\rm b}=56600 M_{\odot}$) and different diffusion coefficients. The red dashed line is the same as in the top panel, while the blue line has a diffusion coefficient that is 10 times larger. Consistent with \citet{Escala_2018}, we find that the simulation predictions are not very sensitive to the choice of diffusion coefficient: the fraction of metal-poor stars that formed before a given $z$ always agrees within $\pm50$\% across the two simulations. We also find that the two simulations' stellar metallicity distributions (Figure~\ref{fig:feh_hist}) are in good agreement at all redshifts. 

\bsp	
\label{lastpage}
\end{document}